\documentclass[aps, reprint, superscriptaddress]{revtex4-2}
\usepackage[utf8]{inputenc}
\usepackage{mathtools}
\usepackage{amsfonts, amsmath, amsthm, amssymb, mathrsfs}
\usepackage{microtype}
\usepackage{graphicx}   
\usepackage{booktabs}
\usepackage[usenames,dvipsnames]{xcolor}    
\usepackage{braket}
\usepackage{tikz}
\usepackage[normalem]{ulem} 
\usepackage{enumitem}                
\usepackage{comment}
\usepackage[hidelinks]{hyperref}
\usepackage{url}
\usepackage[capitalize]{cleveref}
\usepackage{mathtools}
\DeclarePairedDelimiter{\abs}{\lvert}{\rvert}

\crefname{figure}{Fig.}{Figs.}
\crefname{equation}{Eq.}{Eqs.}

\definecolor{myred}{HTML}{e41a1c}
\definecolor{myblue}{HTML}{377eb8}
\definecolor{mygreen}{HTML}{4daf4a}

\hypersetup{
    colorlinks,
    linkcolor={myred!80!black},
    citecolor={MidnightBlue!70!black},
    urlcolor={MidnightBlue!70!black}
}

\newcommand{\tr}{\operatorname{Tr}}
\newcommand{\Qav}{\langle Q_A \rangle}
\newcommand{\rhoa}{\rho_ A}
\newcommand{\Sg}{S^{\mathrm g}}
\newcommand{\Zg}[1]{\mathcal Z^\mathrm g_{#1}}
\newcommand{\rhoaq}{\rho_{A,q}}
\newcommand{\Le}{\ell_\mathrm e}
\newcommand{\Erf}{\operatorname{Erf}}

\newcommand{\sign}{\operatorname{sign}}
\newcommand{\Li}{\operatorname{Li}}
\renewcommand{\Re}{\operatorname{Re}}
\renewcommand{\Im}{\operatorname{Im}}

\begin{document}

\title{Symmetry-resolved entanglement in critical non-Hermitian systems}
\author{Michele Fossati}
\affiliation{SISSA and INFN Sezione di Trieste, via Bonomea 265, 34136 Trieste, Italy} 
\author{Filiberto Ares}
\affiliation{SISSA and INFN Sezione di Trieste, via Bonomea 265, 34136  Trieste, Italy}
\author{Pasquale Calabrese }
\affiliation{SISSA and INFN Sezione di Trieste, via Bonomea 265,   34136 Trieste, Italy} 
\affiliation{International Centre for Theoretical Physics (ICTP), Strada Costiera 11, 34151 Trieste, Italy}

\begin{abstract}
The study of entanglement in the symmetry sectors of a theory has 
recently attracted a lot of attention since it provides better 
understanding of some aspects of quantum many-body systems. In 
this paper, we extend this analysis to the case of non-Hermitian
models, in which the reduced density matrix $\rhoa$ may be non-positive 
definite and the entanglement entropy negative or even complex. 
Here we examine in detail the symmetry-resolved
entanglement in the ground state of the non-Hermitian 
Su-Schrieffer-Heeger chain at the critical point, a model that 
preserves particle number and whose scaling limit is a 
$bc$-ghost non-unitary CFT. By combining bosonization techniques 
in the field theory and exact lattice numerical calculations, we analytically derive the charged moments of $\rhoa$ 
and $|\rhoa|$. From them, we can understand the origin of 
the non-positiveness of $\rhoa$ and naturally define a 
positive-definite reduced density matrix in each charge sector, which
gives a well-defined symmetry-resolved entanglement entropy. As 
byproduct, we also obtain the analytical distribution of the critical entanglement spectrum.
\end{abstract}
\maketitle

\section{Introduction}

Last years have witnessed a growing interest in non-Hermitian 
quantum mechanics. This has been particularly motivated by the 
appearance of non-Hermitian Hamiltonians in the effective
description of a wide variety of phenomena~\cite{Moiseyev, ashida}, as, for example, in the analysis of the $PT$ symmetry~\cite{Bender_1998, Bender_2015, ganainy_2018}, in the study of optical effects~\cite{feng_2017, miri_2019}, or in the investigation of the non-equilibrium properties of open and dissipative systems~\cite{graefe_2008, rotter_2009, muller_2012} as well as measurement-induced transitions~\cite{gopalakrishan_21, biella_21, turkeshi_2021, Muller_2022, turkeshi_2023}, to cite some of them.

A crucial feature of quantum systems is entanglement, not only 
because it is the essential ingredient for performing 
classically-impossible tasks, but also because it has been found
to be a key to understand many physical phenomena in the quantum realm~\cite{afov-08,ccd-09,laflorencie-16,rt-17}.
The main quantity to analyze entanglement in extended quantum systems
is entanglement entropy, which quantifies the amount of entanglement in bipartite
settings. One of its most important properties is its behaviour with 
the size of the subsystem considered~\cite{ecp-10}, as it can be used as an order 
parameter to detect quantum phase transitions and extract valuable 
information about the critical point. In fact, in one-dimensional
Hermitian systems with zero mass gap, the ground state entanglement entropy is
proportional to the central charge of the unitary conformal field theory (CFT) that describes the 
low-energy physics~\cite{Holzhey_94, Calabrese_2004, Calabrese_2009}. 

However, entanglement has been much less studied
in non-Hermitian systems. The majority of the works 
focus on the analysis of the ground state entanglement in critical 
systems described by non-unitary CFTs~\cite{bianchini-15, bianchini-15-2, br-16, narayan, Couvreur_2017, Dupic_2018, Chang_2020, Tu2022GenericEntropy}. 
The main difficulty when studying entanglement in non-Hermitian models
is the lack of a proper entanglement measure: the reduced density matrix can 
be non-positive definite and, consequently, the entanglement entropy can actually 
be negative or even complex. For this reason, several non-Hermitian extensions
of it have been proposed in the literature, which include both changing explicitly 
the definition of the entanglement entropy as in Ref.~\cite{Tu2022GenericEntropy}, or implicitly by introducing 
a modified version of the partial trace as in Ref.~\cite{Couvreur_2017}. A common feature of all
these approaches is that they are sensitive to criticality and provide information about the non-unitary CFT under study.

Recent experiments with cold atoms and ion traps have shown that some 
properties of many-body quantum systems can be understood by 
analyzing the entanglement in the symmetry sectors of the 
theory~\cite{Lukin-19, Azses_2020, Neven_2021, Vitale_2022, Rath_2022}. The basic tool
to investigate how entanglement distributes among symmetry sectors is 
the symmetry-resolved entanglement entropy~\cite{Laflorencie_2014, Goldstein_2018, Xavier_2017}, which provides a finer measure
of the entanglement content in extended systems that is not accessible 
from the total entanglement entropy. This fact has currently triggered an 
intense research activity on the interplay between symmetries and entanglement 
in very different systems, including spin chains~\cite{Xavier_2017, Goldstein_2018, Bonsignori_2019, mdgc-20-2, SREE2dG, ccgm-20, wv-03, bhd-18, bcd-19, byd-20, mrc-20, tr-19, ms-21, amc-22, jones-22, pvcc-22, hms-22,mcp-22}, integrable~\cite{mdgc-20, hc-20, hcc-21, hcc-a-21, chcc-a22, ccadfms-22, ccadfms-22-2, cmca-23, mca-23} and conformal 
field theories~\cite{Goldstein_2018, Xavier_2017, goldstein1, crc-20, MBC-21, Chen-21, Capizzi-Cal-21, Hung-Wong-21, cdm-21, boncal-21, eim-d-21, Chen-22, mt-22, Ghasemi-22, amc-22-2, fmc-22, dgmnsz-22, cmc-23}, and involving diverse contexts such as disorder~\cite{trac-20, kiefer-20, kufs-21-1, kusf-20}, 
non-equilibrium dynamics~\cite{goldstein2, pbc-21-0, pbc-21, fg-21, pbc-22, sh-22, chen-22-2, bcckr-22,amc-22-3,amvc-23}, topological phases~\cite{clss-19, ms-20, Azses-Sela-20, ahn-20, ads-21, ore-21} or holography~\cite{znm-20, wznm-21, znwm-22, bbcg-22}. 

In this paper, we extend the notion of symmetry resolution of entanglement
to the non-Hermitian case and explore if it sheds light on how to better grasp 
entanglement in this kind of systems. To this end, we examine in detail the symmetry resolution 
of entanglement in the ground state of the non-Hermitian version of the Su-Schrieffer-Heeger (SSH)
model at criticality. This is the simplest one-dimensional quadratic fermionic chain that breaks 
Hermiticity with a global $U(1)$ symmetry associated to particle number conservation,
and presents non-trivial features, such as topological phases~\cite{rl-09,liang-13, Lieu_2018, Yao_2018, Hua_2022}. The critical point is 
described by the $bc$-ghost CFT with central charge $c=-2$~\cite{Chang_2020}. Some properties of the critical 
ground state entanglement of this model have been already examined in Refs.~\cite{narayan, Chang_2020, Tu2022GenericEntropy}; 
in particular, it has been found that both the usual entanglement entropy and the 
generalized version introduced in Ref.~\cite{Tu2022GenericEntropy} are proportional to the central charge at 
leading order in the subsystem size and, therefore, they are negative. Away from criticality,
the focus of attention has been the analysis of the entanglement spectrum~\cite{Chang_2020, Herviou_2019}, that is the spectrum
of the reduced density matrix, since it offers useful insights on the topology of the 
model~\cite{Li_2008}.

In Hermitian systems, the entropy of each symmetry sector is usually calculated via the charged moments of the reduced density matrix~\cite{Goldstein_2018}. Inspired by the generalized entanglement entropy defined in Ref.~\cite{Tu2022GenericEntropy}, we consider here the charged moments of the absolute value of the reduced density matrix, which we call \textit{absolute charged moments}. We compute them in the ground state of the $bc$-ghost theory. The analysis of the resulting expression allows us to trace back the origin of the negativeness of the reduced density matrix and, consequently, of the (generalized) R\'enyi entanglement entropies in the models described by this CFT: the sign of the reduced density matrix eigenvalues depends on the parity of the charge sector to which they belong. With this observation, we can define a positive semi-definite density matrix for each charge sector and, from it, a positive symmetry-resolved R\'enyi entanglement entropy. 

Another remarkable consequence of the charge-dependent signature of the entanglement spectrum in the $bc$-ghost theory is that the expression of the standard R\'enyi entanglement entropies depends on the parity of the R\'enyi index $n$. As byproduct of this result, we analytically obtain the distribution of the entanglement spectrum in this CFT, which differs from the unitary case~\cite{CalabreseLefevre}, and we check it against the exact numerical entanglement spectrum of the critical non-Hermitian SSH model.

The paper is organized as follows. In Sec.~\ref{sec:ssh}, we introduce
the non-Hermitian SSH chain and we diagonalize it. In Sec.~\ref{sec:symm_res_ent}, we review the necessary quantities to
generalize the concept of symmetry-resolved entanglement to non-Hermitian systems. In Sec.~\ref{sec:bc_ghost}, we consider the $bc$-ghost CFT and we obtain the absolute charged moments of the ground state reduced density matrix. With them, in Sec.~\ref{sec:symm_res_ent_bc},
we calculate the symmetry-resolved entanglement entropy in this system.
In Sec.~\ref{sec:ordinary_renyi_entropy} we apply the results of the previous section to compute
the standard R\'enyi entanglement entropies, from which, in Sec.~\ref{sec:ent_spec},
we determine the distribution of the entanglement spectrum. We end
in Sec.~\ref{sec:conclusions} with the conclusions. We also include an appendix with
the details of the derivation of the entanglement spectrum distribution.

\section{$PT$ symmetric non-Hermitian SSH model}\label{sec:ssh}

\begin{figure}[t]
\centering
\includegraphics[width=0.67\columnwidth]{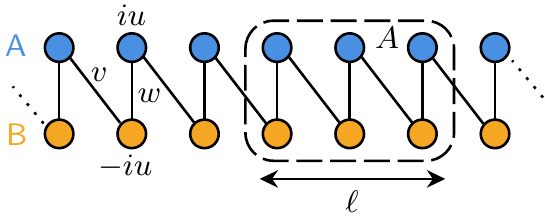}
\caption{Schematic representation of the non-Hermitian SSH model described by the Hamiltonian of \cref{eq:n_H_ssh}. 
 We denote by $v$ and $w$ the hopping amplitudes between the rows $\mathsf A$ and $\mathsf B$ of the ladder. 
 The chemical potential is imaginary, $iu$ for the sites in row $\mathsf A$ and $-iu$ for those in row $\mathsf B$,
 breaking the Hermiticity of the model. We also consider a subsystem $A$ of length $\ell$.}
\label{fig:ssh}
\end{figure}

The non-Hermitian SSH model with $PT$ symmetry is a one dimensional free fermionic chain
with dimerized nearest neighbour couplings, which can be represented as the ladder of Fig.~\ref{fig:ssh}. We label as $\mathsf A$ and $\mathsf B$
the upper and lower rows respectively; therefore, the sites of the ladder form a lattice $\Lambda = \{\mathsf A,\mathsf B\} \times \{1, \dots, L \}$.
We denote by $c_{\sigma j}^\dagger$ and $c_{\sigma j}$  the creation and annihilation fermionic operators 
on the site $(\sigma, j) \in \Lambda$ and by $n_{\sigma j} = c^\dagger_{\sigma j} c_{\sigma j}$ the number operator on that site. 
Then the Hamiltonian of the non-Hermitian SSH model reads \cite{rl-09,liang-13}
\begin{multline}\label{eq:n_H_ssh}
    H= \sum_{j=1}^L \left( -w c_{\mathsf Aj}^{\dagger} c_{\mathsf B j} -v c_{\mathsf A j}^{\dagger} c_{\mathsf B j+1 }+\text { H.c. } \right)  \\
    + i u \sum_{j=1}^L\left(n_{\mathsf Aj}-n_{\mathsf Bj}\right),
\end{multline}
where $w, v$ are positive real hopping parameters, while $iu$, with $u>0$, is a purely imaginary chemical potential, which introduces the non-Hermiticity in the model. We assume periodic boundary conditions $c_{\sigma L+1}\equiv c_{\sigma 1}$.

Observe that the parity transformation acts on the Hamiltonian as $j \mapsto  L+1-j$ 
and exchanging the rows of the ladder, $\mathsf A \leftrightarrow \mathsf B$. Time inversion is 
implemented as the complex conjugation of the parameters. The model has therefore $PT$ symmetry.
Moreover, this Hamiltonian preserves the particle number,
\begin{equation}\label{eq:part_number}
 Q=\sum_{j=1}^L (n_{\mathsf Aj}+n_{\mathsf Bj}),
\end{equation}
i.e. $[H, Q]=0$. This will be the conserved charge with respect to which
we resolve the entanglement in the ground state of the model.

As we mention in the introduction, one of the most relevant 
applications of non-Hermitian quantum Hamiltonians nowadays is to 
measurement-induced phenomena. It has been shown
that the unitary time evolution of a quantum state that is subject to 
a protocol of repeated measurements can be effectively described by a 
non-Hermitian Hamiltonian in the low-rate 
regime of measurements. In particular, in Ref.~\cite{LeGal_2023}, the 
unitary evolution is generated by the Hermitian SSH model ($u=0$ in 
Eq.~\eqref{eq:n_H_ssh}) while the effective dynamics that takes into 
account the effect of the measurements is given by 
Eq.~\eqref{eq:n_H_ssh} 
with $u\neq 0$; in this case, the measurement
protocol yields the staggered imaginary chemical potential of Eq. (1) 
that breaks Hermiticity. An experimental realization of a protocol 
of repeated measurements has been recently reported in 
Ref.~\cite{Koh_2022} using a superconducting quantum processor.

The Hamiltonian~\eqref{eq:n_H_ssh} can be diagonalized as follows. 
After a discrete Fourier transform, 
\begin{equation}\label{eq:fourier_trans}
\tilde{c}_{\sigma k}=\frac{1}{\sqrt{L}} \sum_{j=1}^L e^{i k j} c_{\sigma j}, \quad k=\frac{2 \pi m}{L},
\end{equation}
with $m=0, \ldots, L-1$, the Hamiltonian~\eqref{eq:n_H_ssh} can be cast in the form
\begin{equation}
\begin{aligned}
    H &= \sum_k 
    \tilde{\mathbf{c}}_k^\dagger
    \mathcal H_k
    \tilde{\mathbf{c}}_k,
\end{aligned}
\end{equation}
where $\tilde{\mathbf c}_k^\dagger=(\tilde c_{\mathsf A k}^\dagger, \tilde c_{\mathsf B k}^\dagger)$ and
\begin{equation}
    \mathcal H_k = 
    \begin{pmatrix}
    i u & \eta_k \\
    \eta_k^* & -i u
    \end{pmatrix},
    \qquad
    \eta_k = - w - v e^{-ik}.
\end{equation}
If $u \neq 0$, $\mathcal H_k$ is not Hermitian but it is still diagonalizable for almost every $k$, with an exception that will be discussed later. In fact, there exists an invertible matrix $M(k)$,
\begin{equation}
 M(k)=\left(\begin{array}{cc}
             \frac{\eta_k}{|\eta_k|} \cos(\xi_k) & -\frac{\eta_k}{|\eta_k|}\sin(\xi_k)\\
              \sin(\xi_k) & \cos(\xi_k)
            \end{array}\right),
\end{equation}
where $2 \xi_k = \tan^{-1} ( \abs{\eta_k} / (i u) ) $, such that
\begin{equation}
    \mathcal H_k = M(k) E(k) M(k)^{-1},
\end{equation}
with $E(k) = \mathrm{diag}(\varepsilon_{+,k}, \varepsilon_{-,k})$ and 
eigenvalues
\begin{equation}
 \varepsilon_{\pm, k}=\pm\sqrt{|\eta_k|^2-u^2}.
\end{equation}
Therefore, introducing the new creation and annihilation operators
$\mathbf{f}_k^\dagger=(f_{+,k}^\dagger, f_{-,k}^\dagger)$ and
$\mathbf{d}_k=(d_{+,k}, d_{-,k})^t$, 
\begin{equation}\label{eq:bog_op}
    \mathbf f^\dagger_k = \tilde{\mathbf c}^\dagger_k M(k) \qquad \mathbf d_k = M(k)^{-1} \tilde{\mathbf c}_k,
\end{equation}
the Hamiltonian~\eqref{eq:n_H_ssh} is diagonal in terms of them,
\begin{equation}
    H = \sum_{k} \mathbf f^\dagger_k E(k) \mathbf d_k.
\end{equation}
The model presents three different phases~\cite{Lieu_2018}. If $w - v > u$, the system is in a $PT$-unbroken phase, with real spectrum and trivial topological properties. If $-u<w - v<u$, the system is in a $PT$-broken phase, with purely imaginary spectrum. If $w - v < -u$, the system is in a $PT$-unbroken phase, with real spectrum and non-trivial topological properties. 
These phases are separated by two critical lines defined by 
$w-v=\pm u$. Along these lines, the spectrum is $\varepsilon_{\pm,k}= \pm \sqrt{2vw(1+\cos k)}$.
In this case, the two bands approach $0$ as $k\to\pi$, and the eigenspaces tend to become
collinear so that $\mathcal{H}_\pi$ is not diagonalizable: $k=\pi$ is commonly referred
to as an \textit{exceptional point}.

The eigenstates of the Hamiltonian~\eqref{eq:n_H_ssh} can be constructed in the following way. Let us introduce
the set of labels of the single particle eigenstates of $H$, $\Gamma=\{+,-\}\times \{k\}$.
The vacuum state $\ket 0$ is defined as the one annihilated by $d_p$ for every $p \in \Gamma$ and, therefore, $H \ket 0 = 0$. 
The \textit{right} eigenstates of $H$ are then obtained by applying sequences of $f^\dagger_p$ operators to the vacuum. More precisely, let $B$ be a subset of $\Gamma$, then $\ket{B_R} = \prod_{p \in B} f^\dagger_p \ket 0$ is a right eigenstate of $H$ with eigenvalue $\sum_{p \in B} \varepsilon_p$.
Analogously, \textit{left} eigenstates are constructed by applying sequences of $d_p$ operators to $\bra 0$ 
on the right, thus $\bra{B_L} = \bra 0 \prod_{p \in B} d_p$ is a left eigenstate of $H$  with eigenvalue 
$\sum_{p \in B} \varepsilon_p$.

If the spectrum of $H$ is real, then the eigenvalues can be ordered and the notion of ground state is well defined: it is the many-body state in which all the single particle states with negative energy are occupied. We call $\ket{\psi_R}$ the right ground state and $\bra{\psi_L}$ the left ground state, which are explicitly 
\begin{equation}\label{eq:g_s_ssh}
    \bra{\psi_L} = \bra 0 \prod_{k} d_{k,-}  
    \qquad 
    \ket {\psi_R} = \prod_{k} f^\dagger_{k,-} \ket 0.
\end{equation}
In a non-Hermitian setting, the natural way to define the expectation value of an observable $\mathcal O$ in the 
ground state of $H$  is $\bra{\psi_L} \mathcal O \ket{\psi_R}$, see Ref.~\cite{Brody_2014}. It follows then that 
the ground state density matrix is $\rho = \ket{\psi_R} \bra{\psi_L}$.
Observe that, if the system is at temperature $1/\beta$, its state may be described by the Gibbs ensemble $\rho_\beta=e^{-\beta H}/{\rm Tr}(e^{-\beta H})$, which in the zero-temperature limit leads
to $\rho=\ket{\psi_R}\bra{\psi_L}$. As shown in detail in Refs.~\cite{bianchini-15, Dupic_2018}, this fact facilitates the study in field theories of the entanglement entropies of $\rho$ since it allows, within the path integral formalism, to straightforwardly identify the moments of the reduced density matrix with partition functions on a replicated Riemann surface or, alternatively, with correlators of twist fields.

One may consider other density matrices, such as $\ket{\psi_R}\bra{\psi_R}$, which is Hermitian, and yields a positive definite reduced density matrix and entanglement entropies. Some entanglement
properties of this density matrix in the gapped regions of the Hamiltonian~\eqref{eq:n_H_ssh} have been examined in Ref.~\cite{Herviou_2019}. However, as observed in Ref.~\cite{Dupic_2018}, the R\'enyi entanglement entropies are harder
to compute in field theory since the connection with partition functions on a replicated surface is not as direct as in the case 
$\ket{\psi_R}\bra{\psi_L}$, it involves a time reversal operation that is in general a difficult problem. Since we are interested in study the universal properties of the symmetry-resolved ground state 
entanglement of the critical non-Hermitian SSH chain and the corresponding CFT, in the rest of the paper we take as ground state the density matrix $\rho=\ket{\psi_R}\bra{\psi_L}$. We refer the reader to
Ref.~\cite{Dupic_2018} for a thorough discussion on the possible choices of the total density matrix in non-Hermitian systems.

A key object in our analysis of the entanglement properties of the 
ground state is the two-point correlation matrix $C$ with entries
\begin{equation}\label{eq:correl_matrix}
    C_{j, j'}=\bra{\psi_L} \mathbf c^\dagger_j \mathbf c_{j'} \ket{\psi_R},
\end{equation}
where here $\mathbf c_j =(c_{\mathsf A j}, c_{\mathsf B j})$. Using Eqs.~\eqref{eq:fourier_trans} and 
\eqref{eq:bog_op} to express the operators 
$\mathbf{c}_j^\dagger$ and $\mathbf{c}_j$ in terms of $\mathbf{f}_k^\dagger$ and $\mathbf{d}_k$, 
one finds
\begin{equation}\label{eq:correl_matrix_2}
    C_{jj'}=  
    \frac{1}{L} \sum_{k} e^{-ik(j-j')} \mathcal{G}(k),
\end{equation}
where
\begin{equation}
    \mathcal{G}(k)=\frac{1}{2}\left(
    \begin{array}{cc}
        1-\cos(2\xi_k) & -\sqrt{\frac{\eta_k^*}{\eta_k}}\sin(2\xi_k) \\
        -\sqrt{\frac{\eta_k}{\eta_k^*}}\sin(2\xi_k) & 1+\cos(2\xi_k)
    \end{array}
    \right).
\end{equation}
Note that, due to the dimerization of the hopping amplitudes $v$, $w$, the correlation matrix $C$ presents a $2\times 2$ block structure. In the thermodynamic limit $L\to\infty$, $C$ is a block Toeplitz matrix generated by the symbol $\mathcal{G}$.
 
At the critical points $w-v = \pm u$, $\mathcal G(k)$ has a singularity at the mode $k = \pi$ and all its entries diverge as $1/\abs{k - \pi}$. Hence, the correlation matrix $C$ is not well-defined. Following Refs.~\cite{Chang_2020, Tu2022GenericEntropy},
a way to deal with this divergence is to perform in Eq.~\eqref{eq:correl_matrix_2} a small shift in the moments $k\mapsto k+\kappa/L$, with $\kappa\ll 1$. 
The numerical data presented in this work has been obtained taking $\kappa=10^{-7}$. 

As we have pointed out in the introduction, in this paper
we are interested in studying the symmetry resolution of the
ground state entanglement with respect to the $U(1)$ symmetry associated with particle number conservation at criticality.
Therefore, in the following, we will restrict our analysis to the critical line $u = w - v$, which separates the $PT$-unbroken and topologically trivial phase from the $PT$-broken phase.

\section{Symmetry-resolved entanglement}\label{sec:symm_res_ent}
In this section, we introduce all the quantities that we will use to 
investigate the symmetry-resolved entanglement in the ground state of 
the critical non-Hermitian SSH model previously discussed. 

We consider a spatial bipartition of the system $A \cup \bar{A}$ with $A$ a subset of contiguous sites of length $\ell$ as the one depicted in Fig.~\ref{fig:ssh}. Then the total
Hilbert space factorizes into $\mathscr H_A \otimes \mathscr H_{\bar A}$. 
As we have seen in the previous section, the ground state density matrix is $\rho = \ket{\psi_R}\bra{\psi_L}$.
The reduced density matrix that describes the state of subsystem $ A$ is obtained
by taking the partial trace to the complementary subsystem $\bar{A}$, $\rhoa = \tr_{\bar{A}}(\rho)$. 

In a non-Hermitian system, we have to take into account that $\rho$ is positive semi-definite, $\rho \geq 0$, but is generally not Hermitian, $\rho^\dagger\neq \rho$ \cite{Brody_2014}. 
This fact implies that $\rho_A$ may not be positive semi-definite. Thus the eigenvalues of $\rhoa$ can be negative or even complex.
In any case, $\tr(\rho)=\tr(\rho_{A})=1$.

The entanglement entropy of the bipartition $A \cup \bar{A}$ is defined as 
\begin{equation}\label{eq:vn_ent_ent}
S(\rhoa) = - \tr(\rhoa \log \rhoa).
\end{equation}
For Hermitian systems, $S(\rhoa)$ measures the degree of entanglement 
between subsystems $ A$ and $\bar{ A}$. In particular, if the 
low-energy physics of the model is described by a (unitary) CFT, the entanglement entropy behaves in the thermodynamic limit as 
\begin{equation} \label{eq:c/3logl}
    S(\rhoa) = \frac{c}{3} \log \ell + O(\ell^0)
\end{equation} 
where $c$ is the central charge of the corresponding CFT \cite{Holzhey_94, Calabrese_2004,Calabrese_2009}.

A related family of quantities are the R\'enyi entanglement entropies 
\begin{equation}
S_n(\rhoa) = \frac{1}{1-n} \log \tr(\rhoa^n),
\end{equation}
from which is possible to recover the entanglement entropy of \cref{eq:vn_ent_ent}
through the limit $S = \lim_{n \to 1} S_n$. The scaling behaviour of the R\'enyi entanglement entropies for critical infinite Hermitian 
systems is 
\begin{equation}\label{eq:renyi_ent_unitary_cft}
S_n(\rhoa) = \frac{c}{6}\frac{n+1}{n}\log \ell + O(\ell^0).
\end{equation}

On the other hand, in non-Hermitian systems, since $\rhoa$ is not positive semi-definite, 
the entanglement entropy~\eqref{eq:vn_ent_ent} can be complex and ambiguous, depending on 
which branch of the logarithm we take. In the recent Ref.~\cite{Tu2022GenericEntropy} 
a new quantity dubbed \emph{generalized entanglement entropy} is introduced,
\begin{equation}\label{eq:gen_ent_ent}
    S^{\rm g}(\rhoa) = - \tr(\rhoa \log \abs{\rhoa}).
\end{equation}
Observe that the only difference with the usual entropy~\eqref{eq:vn_ent_ent} is 
that it takes the logarithm of $|\rhoa|$ instead of $\rhoa$:
if $\rhoa$ is diagonalized by the matrix $R$ such that 
$\rhoa=R^{-1}{\rm diag}(\lambda_j)R$, then $|\rhoa| \coloneqq R^{-1}{\rm diag}(|\lambda_j|)R$. This guarantees that $\Sg$ is real when the eigenvalues of $\rhoa$ are either negative or complex in conjugate pairs. Note that, when $\rhoa \geq 0$, it reduces to the standard entanglement entropy, 
$\Sg (\rhoa) = S(\rhoa)$. Although $S^{\rm g}(\rhoa)$ can be negative, 
and its interpretation as an entanglement measure is not clear, it 
satisfies that $S^{\rm g}(\rhoa) \neq 0$ only if the subsystems $A$ and $\bar{A}$ 
are entangled. Moreover, analogously to the standard entanglement entropy, 
it may be useful to extract information about the (non-unitary) CFTs that 
describe critical non-Hermitian systems. In fact, in Ref.~\cite{Tu2022GenericEntropy}, it has been 
verified for a set of critical non-Hermitian models, including the non-Hermitian SSH, 
that $S^{\rm g}(\rhoa)$ scales as the R.H.S. of \cref{eq:c/3logl}, even 
when the central charge of the associated CFT is negative. In Ref.~\cite{Cipolloni_23}, the generalized
entanglement entropy~\eqref{eq:gen_ent_ent} has been employed to study the 
entanglement of typical eigenstates of non-Hermitian systems and get a 
non-Hermitian analogue of the Page curve.

Together with the generalized entanglement entropy, one can introduce 
the \emph{generalized Renyi entropies} 
\begin{equation}\label{eq:gen_renyi_ent}
    S_n^\mathrm g(\rhoa) = \frac{1}{1-n} \log \tr \left( \rhoa \abs{\rhoa}^{n-1} \right)
\end{equation}
that satisfy $S^g = \lim_{n \to 1} S_n^\mathrm g$.

Using the generalized entanglement entropy~\eqref{eq:gen_ent_ent}, we can extend the 
notion of symmetry-resolved entanglement \cite{Laflorencie_2014, Goldstein_2018, Xavier_2017} 
to non-Hermitian systems. Let us assume that the system has an internal $U(1)$ symmetry generated 
by a charge operator $Q$, which in our case will be the particle
number operator of \cref{eq:part_number}. This operator can be decomposed 
as the sum of the charge in $A$ and $\bar{A}$, $Q=Q_A+Q_{\bar{A}}$. If the 
left and the right ground states $\ket{\psi_R}$ and $\bra{\psi_L}$
are eigenstates of $Q$, as occurs in the non-Hermitian SSH model, 
then $\rho = \ket{\psi_R}\bra{\psi_L}$ commutes with $Q$. This implies 
that $\rhoa$ acts separately on each eigenspace of $Q_A$. Let us denote by
$\Pi_q$ the projector onto the eigenspace of $Q_A$ with eigenvalue $q\in\mathbb{Z}$, 
also called \textit{charge sector}. We then can construct a density matrix
$\rhoaq$ for each charge sector,
\begin{equation}
 \rhoaq=\frac{\Pi_q \rhoa\Pi_q}{p(q)},
\end{equation}
where $p(q)=\tr(\Pi_q\rhoa)$ normalizes it to $\tr(\rhoaq)=1$, such that
\begin{equation}\label{eq:charge_dec_rhoa}
    \rhoa =  \bigoplus_q p(q) \rhoaq.
\end{equation}
In Hermitian systems, $p(q)$ is interpreted as the probability of finding the 
subsystem $A$ in the sector with charge $q$. However, now $\rhoa$ is not positive 
definite, so $p(q)$ may be negative, as well as larger than $1$. Hence the 
interpretation of $p(q)$ as a probability is lost in the non-Hermitian case. 

Applying \cref{eq:charge_dec_rhoa}, the generalized entanglement entropy~\eqref{eq:gen_ent_ent} admits 
the following decomposition in the charge sectors of $Q_A$,
\begin{equation} \label{eq:symmetry-res-decomposition}
    \Sg (\rhoa) = - \sum_q p(q) \log \abs{p(q)} + \sum_q p(q) \Sg (\rhoaq).
\end{equation}
In the Hermitian case, the first term on the right hand side is commonly referred 
as the \emph{number entropy} while the second is called 
\emph{configurational entropy}~\cite{bhd-18, bcd-19, Lukin-19}. 

The symmetry-resolved R\'enyi entanglement entropies $\Sg(\rhoaq)$ can be calculated 
using the Fourier representation of the projector $\Pi_q$. In fact, introducing the 
\textit{absolute charged moments} of $\rhoa$
\begin{equation}\label{eq:abs_charged_mom}
    Z^\mathrm g_n(\alpha) = \tr \left( \rhoa \abs{\rhoa}^{n-1} e^{i \alpha Q_A} \right)
\end{equation}
and their Fourier transform
\begin{equation}\label{eq:f_trans_abs_charged_mom}
 \Zg{n} (q) \coloneqq \tr \left(\Pi_q\rhoa \abs\rhoa^{n-1} \right)= \int_{-\pi}^\pi \frac{d \alpha}{2\pi} e^{-i \alpha q} Z^\mathrm g_n (\alpha),
\end{equation}
it is possible to express all the ingredients of \cref{eq:symmetry-res-decomposition} 
in the form $p(q) = \Zg{1} (q)$ and
\begin{equation} \label{eq:def-Sgnq}
    \Sg_n(\rhoaq) = \frac{1}{1-n} \log \frac{\Zg{n}(q)}{\Zg{1}(q) \abs{\Zg{1}(q)}^{n-1}}.
\end{equation}
When $\rhoa$ is positive definite, $Z_n^{\rm g}(\alpha)$ is 
equal to the usual \emph{charged moments} of $\rho_A$,
\begin{equation}\label{eq:standard_charged_moments}
    Z_n(\alpha) = \tr \left( \rhoa^n e^{i\alpha Q_A} \right).
\end{equation}

In quadratic fermionic chains, the moments $Z_n(\alpha), Z_n^\mathrm g (\alpha)$ as well the whole reduced density matrix $\rhoa$
are accessible through the two-point correlation matrix of \cref{eq:correl_matrix} restricted to 
the subsystem $A$. This result was initially derived for Hermitian systems~\cite{Peschel_2003} 
but it also holds in the non-Hermitian case \cite{Herviou_2019}. 
Since the Hamiltonian is quadratic, then the ground state reduced density matrix satisfies the Wick theorem and, therefore, it is in general a Gaussian operator of the form  $\rhoa = e^{-H^\mathrm E}/\mathcal N$, with $H^\mathrm E = \sum_{ij} c^\dagger_i \mathcal H^\mathrm E_{ij} c_j$. The relation with the two-point correlation matrix is set by 
\begin{equation}\label{eq:corr_matrix_ent_ham}
 H^\mathrm E = \log\left[(I-C_A^t)/C_A^t\right]
\end{equation}
and $\mathcal N^{-1} = \det (I-C_A)$, where $C_A$ is the restriction of $C$ to subsystem $A$, and the superscript $t$ indicates transposition. Denoting by $\nu_j$ the eigenvalues of $C_A$, it follows that
\begin{equation}\label{eq:abs_charged_mom_corr_matrix}
     \log Z^\mathrm g_n(\alpha) = \sum_{j=1}^{2\ell} 
     \log \left[\nu_j  |\nu_j|^{n-1} e^{i \alpha} + (1-\nu_j) |1-\nu_j|^{n-1} \right]
\end{equation}
and
\begin{equation}
    \Sg(\rhoa) = - \sum_{j=1}^{2\ell}  \left[ \nu_j \log |\nu_j| + (1-\nu_j) \log |1-\nu_j| \right].
\end{equation}
The standard quantities $Z_n(\alpha)$ and $S(\rhoa)$ follow  
analogue formulae without the absolute values. In particular,
\begin{equation}\label{eq:standard_moments_corr_matrix}
     \log Z_n(\alpha) = \sum_{j=1}^{2\ell} 
     \log \left[\nu_j^n e^{i \alpha} + (1-\nu_j)^n \right].
\end{equation}

\section{Absolute charged moments: the $bc$-ghost CFT}\label{sec:bc_ghost}

In this section, we obtain the absolute charged
moments $Z_n^{\rm g}(\alpha)$, defined in \cref{eq:abs_charged_mom}, in the ground
state of the non-Hermitian SSH model along the critical
line $u=w-v$. To this end, we study them both numerically 
using Eq.~\eqref{eq:abs_charged_mom_corr_matrix} and analytically by considering the non-unitary CFT that 
describes the model at criticality.

On the critical line $u=w-v$, the low-energy physics of the non-Hermitian SSH model of \cref{eq:n_H_ssh} 
is captured by the $bc$-ghost CFT with central charge 
$c=-2$, see Ref.~\cite{Chang_2020}. The action of this theory 
reads~\cite{friedan-86, DiFrancesco}
\begin{equation}
S_{bc}=\int d^2 z(\psi_b\bar{\partial}\psi_c+
\bar{\psi}_b\partial\bar{\psi}_c),
\end{equation}
where $\psi_b(z)$ and $\psi_c(z)$ are anti-commuting 
holomorphic primary fields with conformal weights $h_b=1$
and $h_c=0$ respectively, while $\bar{\psi}_b(\bar{z})$ and
$\bar{\psi}_c(\bar{z})$ denote their anti-holomorphic 
counterparts. 

The $bc$-ghost CFT has a global $U(1)$ symmetry associated 
to \textit{ghost number} conservation, which on the lattice
corresponds to the particle number of \cref{eq:part_number}. 
The holomorphic Noether current is
\begin{equation}\label{eq:ghost_current}
j(z)=-:\psi_b(z)\psi_c(z):.
\end{equation}

To determine the universal terms of the different moments of 
$\rho_A$ introduced in Sec.~\ref{sec:symm_res_ent}, we consider the orbifold theory ${\rm CFT}^{\otimes n}/\mathbb{Z}_n$, obtained by taking $n$ copies on the complex
plane of the CFT under study and then quotient the resulting tensor 
product by the $\mathbb{Z}_n$ symmetry related to the the cyclic permutation of the replicas. 

Before proceeding, it is important to remark that we are assuming that the CFT is defined in the complex plane, which corresponds to taking
an infinite system, i.e. $L\to \infty$, while the numerical 
calculations are done for a finite chain with periodic boundary 
conditions. In CFT, a finite periodic system is obtained by 
conformally mapping the complex plane to a cylinder. Since the moments
of $\rhoa$ are given by correlators of primary fields, the effect of this map in the expression for the moments of $\rhoa$ is simply to replace the subsystem length $\ell$ by the chord length
\begin{equation}
    \Le = \frac{2 L}{\pi} \sin \left( \frac{\pi \ell}{L} \right).
\end{equation}
The factor 2 is due to the number of sites that in the full SSH chain and in the subsystem $A$ is $2L$ and $2\ell$ respectively. 

In unitary CFTs, the ground state neutral moments $Z_n(0)=\tr(\rho_A^n)$ 
are given by the two-point correlation function of the orbifold~\cite{Calabrese_2004, Calabrese_2009, ccad-07}
\begin{equation}\label{eq:unitary_twist_corr}
    Z_n(0)=\bra{0} \tau_n(0) \tilde{\tau}_n(\ell) \ket{0}.
\end{equation}
 Here $\ket{0}$ denotes the conformal vacuum, i.e.
the state of the CFT invariant under global conformal transformations,
and the fields $\tau_n$, $\tilde{\tau}_n$ are called twist and anti-twist fields~\cite{dixon-87, k-87}. 
The winding around the point where $\tau_n$ ($\tilde{\tau}_n$) is inserted takes a field $\mathcal{O}_k$ living 
in the copy $k$ of the orbifold into the copy $k+1$ ($k-1$), for example
\begin{equation}
    \mathcal{O}_k(e^{2\pi i}z)\tau_n(0)=\tau_n(0)\mathcal{O}_{k+1}(z).
\end{equation}
 The twist and anti-twist fields are spinless primaries with conformal
 dimension
 \begin{equation}
     h_n^{\tau}=\frac{c}{24}\left(n-\frac{1}{n}\right)
 \end{equation}
 and, therefore, \cref{eq:unitary_twist_corr} directly gives \cref{eq:renyi_ent_unitary_cft} for the ground state R\'enyi entanglement 
 entropy of Hermitian systems at the critical point.

In non-unitary CFTs, the previous discussion is more subtle because 
the physical ground state is not the conformal vacuum $\ket{0}$. In
general, the physical ground state corresponds to the lowest eigenstate
of the Virasoro operator $L_0$. In unitary CFTs, such eigenstate is the 
conformal vacuum since all the non-trivial primary fields have positive 
conformal dimension. On the contrary, in non-unitary CFTs, we can find 
primary fields with negative dimension. This implies that the ground 
state is the state $\ket{\phi}=\phi(0)\ket{0}$ associated to the primary 
$\phi$ with the lowest dimension $h_\phi\leq 0$. In Refs.~\cite{bianchini-15, bianchini-15-2}, 
it was proposed that in such case the ground state neutral moments are given 
by the orbifold two-point correlations
\begin{equation}\label{eq:non_unitary_comp_twist}
    Z_n(0)=\frac{\bra{0}\tau_n^{\phi}(0)\tilde{\tau}_n^{\phi}(\ell)\ket{0}}{\bra{0}\phi(0)\phi(\ell)\ket{0}^n},
\end{equation}
 involving the composite field $\tau_n^{\phi}\equiv \tau_n\cdot \phi$,
 which is a spinless primary field of the orbifold with dimension
 \begin{equation}
     h_n^{\tau^{\phi}}=h_n^{\tau}+\frac{h_\phi}{n}.
 \end{equation}
Therefore, \cref{eq:non_unitary_comp_twist} implies the following behaviour 
for $Z_n(0)$ in terms of the subsystem size $\ell$
\begin{equation}\label{eq:non_unitary_comp_twist_2}
 Z_n(0)\sim\ell^{-\frac{c_{\rm eff}}{6}\left(n-\frac{1}{n}\right)},
\end{equation}
 where $c_{\rm eff}=c-24h_\phi$. In the case of the $bc$-ghost theory,
 the lowest dimension field $\phi$ is $\psi_c$, which has dimension
 $h_c=0$, and then one may conclude that, in this case, $c_{\rm eff}=c=-2$. 
 
\begin{figure}
     \centering
         \includegraphics[width=\columnwidth]{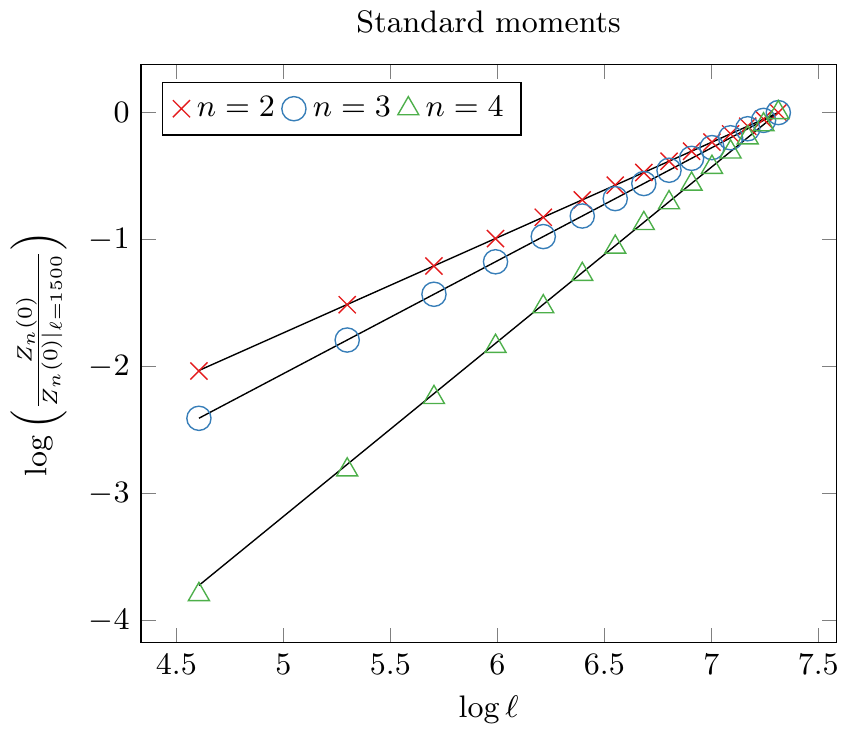}
         \hspace{0.2cm}
         \centering
         
         \includegraphics[width=\columnwidth]{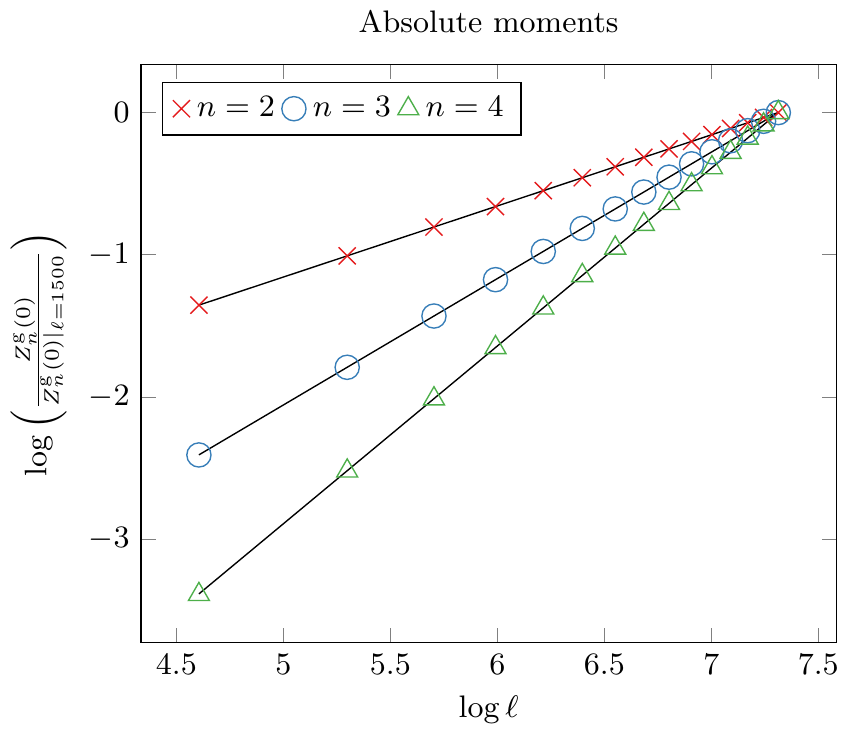}
     
     \caption{Standard moments $Z_n(0)$ (upper panel) and 
     absolute moments $Z_n^{\rm g}(0)$ (lower panel) of $\rhoa$ as a function
     of the subsystem size $\ell$ and different values of the exponent $n$. We plot the ratio with the corresponding value for $\ell=1500$, taken as an arbitrary reference point. The symbols are their exact numerical value for the non-Hermitian SSH chain of length $L=\ell/2$ and parameters $w=3/2$, $v=1$ and $u=1/2$, computed 
     through the two-point correlation matrix~\eqref{eq:correl_matrix_2} using Eq.~\eqref{eq:standard_moments_corr_matrix}
     for $Z_n(0)$ and Eq.~\eqref{eq:abs_charged_mom_corr_matrix} for $Z_n^{\rm g}(0)$. The solid lines
     correspond to the analytic expressions~\eqref{eq:standard_moments_ssh} in the upper panel
     and ~\eqref{eq:abs_neutral_mom_ssh} in the lower one, upon replacing $\ell$ by the chord
     length $\ell_{\rm e}$.} 
     \label{fig:neutral_moments}
\end{figure}

 However, the numerical analysis of the moments $Z_n(0)$ in the ground state of the non-Hermitian SSH model 
 reveals that they do not behave as expected from \cref{eq:non_unitary_comp_twist_2} for any integer
 $n$ but as
 \begin{equation}\label{eq:standard_moments_ssh}
  \log Z_n(0)=\left\{
  \begin{array}{ll}
   \frac{1}{3}\left(n-\frac{1}{n}\right)\log \ell + O(\ell^0) & n \text{ odd},\\
   \\
   \left(\frac{n}{3}+\frac{1}{6n}\right)\log \ell + O(\ell^0) & n \text{ even}.
  \end{array}\right.
 \end{equation}
In fact, in the upper panel of \cref{fig:neutral_moments}, we numerically study $\log Z_n(0)$ as a function of the subsystem
length, taking $\ell=L/2$ and varying the total system size for
several values of $n$.
The points are the exact numerical value of $\log Z_n(0)$ for the non-
Hermitian SSH model obtained by diagonalizing the correlation matrix~\eqref{eq:correl_matrix} and applying \cref{eq:standard_moments_corr_matrix}. The continuous lines correspond to \cref{eq:standard_moments_ssh}.

On the other hand, as the lower panel of Fig.~\ref{fig:neutral_moments} shows, we find that the absolute moments $Z_n^{\rm g}(0)={\rm Tr}(\rho_A|\rho_A|^{n-1})$ do follow the behaviour of
 \cref{eq:non_unitary_comp_twist_2}  for any $n$,
 \begin{equation}\label{eq:abs_neutral_mom_ssh}
  \log Z_n^{\rm g}(0)=\frac{1}{3}\left(n-\frac{1}{n}\right)\log\ell+O(\ell^0),
 \end{equation}
 as also has been numerically checked in Ref.~\cite{Tu2022GenericEntropy}. 
 In Fig.~\ref{fig:neutral_moments}, the points are the exact numerical value of $\log Z_n^{\rm g}(0)$
 taking as subsystem length $\ell=L/2$ and varying $L$ for several R\'enyi indices $n$. 
 They have been calculated using the two-point correlation matrix~\eqref{eq:correl_matrix} 
 through \cref{eq:abs_charged_mom_corr_matrix}. The continuous lines correspond
 to \cref{eq:abs_neutral_mom_ssh}.

 The previous analysis leads to conclude that, in the case of the $bc$-ghost theory, 
 the ground state absolute moments of $\rhoa$ are given by the orbifold correlation
 function
 \begin{equation}\label{eq:abs_mom_cft}
     Z_n^{\rm g}(0)=\bra{0}\tau_n(0)\tilde{\tau}_n(\ell)\ket{0}.
 \end{equation}
In Sec.~\ref{sec:ordinary_renyi_entropy}, we will see that the 
alternating behavior in $n$ of the standard moments 
in~\cref{eq:standard_moments_ssh} originates from the property that the eigenvalues of $\rhoa$ have constant 
sign in each charge sector and this sign depends on the parity of the 
charge. Combining this property and Eq.~\eqref{eq:abs_mom_cft}, we 
will determine the twist field correlation function that gives the 
correct result for $n$ even in Eq.~\eqref{eq:standard_moments_ssh}.

The absolute charged moments $Z_n^{\rm g}(\alpha)$ of \cref{eq:abs_charged_mom} 
can also be calculated using the orbifold ${\rm CFT}^{\otimes n}/\mathbb{Z}_n$. In the orbifold, the operator
$e^{i\alpha Q_A}$ has the effect of adding a phase $e^{i\alpha}$ when
a charged particle moves along a closed path that crossed all the copies. This phase shift can be incorporated in the correlators by 
inserting a local $U(1)$ operator $\mathcal{V}_\alpha(z)$ that gives
rise to a term $e^{i\alpha}$ when going around it. Similarly to unitary
CFTs with a $U(1)$ symmetry, see Ref.~\cite{Goldstein_2018}, we can construct 
the composite twist field $\tau_{n,\alpha}\equiv \tau_n\cdot \mathcal{V}_\alpha$ such that, when
winding around the point where it is inserted, a field $\mathcal{O}_k$
living in the copy $k$ is mapped to the copy $k+1$ with an extra phase
$e^{i\alpha/n}$, 
\begin{equation}
 \mathcal{O}_k(e^{2\pi i}z)\tau_{n,\alpha}(0)=
 e^{i\alpha/n} \tau_{n,\alpha}(0)\mathcal{O}_{k+1}(z).
\end{equation}
If $\mathcal{V}_\alpha$ is a spinless primary field with conformal
dimension $h_\alpha^{\mathcal{V}}$, then the composite twist field
$\tau_{n,\alpha}$ is also a spinless primary of dimension~\cite{Goldstein_2018}
\begin{equation}
h_{n,\alpha}=h_n^{\tau}+\frac{h_\alpha^{\mathcal{V}}}{n}.
\end{equation}
Thus we propose that, in the $bc$-ghost theory, the absolute
charged moments are given by the orbifold two-point correlation
function
\begin{equation}\label{eq:abs_charged_mom_bc}
    Z_n^{\rm g}(\alpha)=
    \bra{0} \tau_{n,\alpha}(0)\tilde{\tau}_{n,-\alpha}(\ell)\ket{0}.
\end{equation}
Observe that this expression simplifies to \cref{eq:abs_mom_cft} 
when $\alpha=0$. 

The field $\mathcal{V}_\alpha$ can be determined by applying the fact 
that the $bc$-ghost model is equivalent via bosonization to a Gaussian
theory coupled to a background charge~\cite{friedan-86}. The bosonization prescription
works as follows. Let us consider a free bosonic field theory and we introduce a background charge $\mathcal{Q}$,
\begin{equation}\label{eq:gaussian_background_action}
S_{\mathcal{Q}}=\frac{1}{2\pi}\int d^2z \left(\partial \varphi \bar{\partial}\varphi+\frac{i \mathcal{Q}}{4} \varphi\right).
\end{equation}
 The scalar field $\varphi(z)$ is compactified on a circle of radius one, i.e. $\varphi\sim \varphi +2\pi m$ with $m\in\mathbb{Z}$.
The inclusion of a background charge in the action~\eqref{eq:gaussian_background_action} modifies the 
central charge of the theory to $c=1-3\mathcal{Q}^2$ as well as 
the conformal dimension of the vertex operators $V_\alpha(z)=e^{i\sqrt{2}\alpha \varphi(z)}$, which is now $h_\alpha^V=\alpha^2-\alpha \mathcal{Q}/\sqrt{2}$. On the other hand, in presence of a background charge, 
$\partial \varphi$ is not a primary. If we take $\mathcal{Q}=1$, 
this theory has central charge $c=-2$, 
the fields $\psi_b$ and $\psi_c$ are identified with the vertex operators
\begin{equation}
\psi_b(z)\sim V_{-\frac{1}{\sqrt{2}}}(z),\quad 
\psi_c(z)\sim V_{\frac{1}{\sqrt{2}}}(z),
\end{equation}
and the $U(1)$ current 
\begin{equation}
    j(z)=-\partial \varphi(z)
\end{equation}
is equivalent to the ghost current of \cref{eq:ghost_current}. 

Therefore, using these bosonization relations, the charge operator
in the interval $A$ is
\begin{equation}
    Q_A=-\frac{1}{2\pi}\int_A dx \partial_x \varphi=\frac{1}{2\pi}\left(\varphi(0)-\varphi(\ell)\right),
\end{equation}
and the local $U(1)$ operator $\mathcal{V}_\alpha$ can be identified
as the vertex operator
\begin{equation}
    \mathcal{V}_\alpha(z)= e^{i \frac{\alpha}{2\pi}\varphi(z)}
\end{equation}
with conformal dimension
\begin{equation}\label{eq:conf_dim_vertex_bc}
    h_\alpha^{\mathcal{V}}=\frac{|\alpha|}{4\pi}\left(\frac{|\alpha|}{2\pi}-1\right).
\end{equation}
Notice that such dimension is different from the standard one in the Hermitian CFT which is proportional to $\alpha^2$. 

Finally, applying Eq. \eqref{eq:conf_dim_vertex_bc} in Eq. \eqref{eq:abs_charged_mom_bc},  we conclude that the ground state absolute charged moments of the 
non-Hermitian SSH model along the critical line are of the form 
\begin{equation}\label{eq:final_abs_charged_mom}
 Z_n^{\rm g}(\alpha)=c_{n,\alpha}e^{i\alpha\langle Q_A\rangle}\ell^{\frac{1}{3}\left(n-\frac{1}{n}\right)
 -\frac{|\alpha|}{\pi n}\left(\frac{|\alpha|}{2\pi}-1\right)}.
\end{equation}
Observe that in this expression we have included a phase 
$e^{i\alpha\langle Q_A\rangle}$, which depends on the average 
charge $\langle Q_A\rangle$ in the subsystem $A$. This term
is not captured by the previous CFT calculations since it depends 
on the particular microscopical model under consideration. In the
critical non-Hermitian SSH chain, the ground state is half-filled, 
see \cref{eq:g_s_ssh}, and $\langle Q_{A}\rangle=\ell$.

\begin{figure}
    \centering
    \includegraphics[width=\columnwidth]{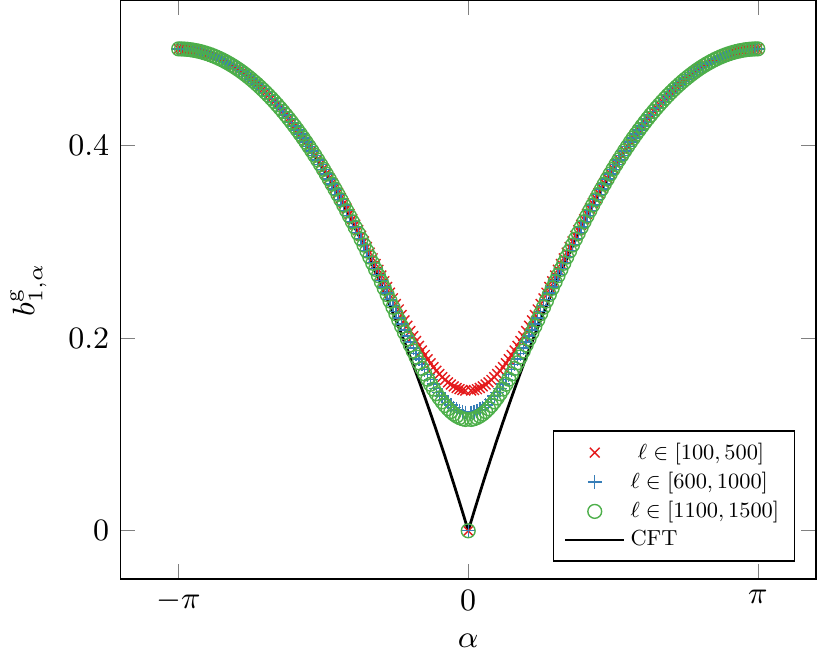}

         \hspace{0.2cm}
    \centering
    \includegraphics[width=\columnwidth]{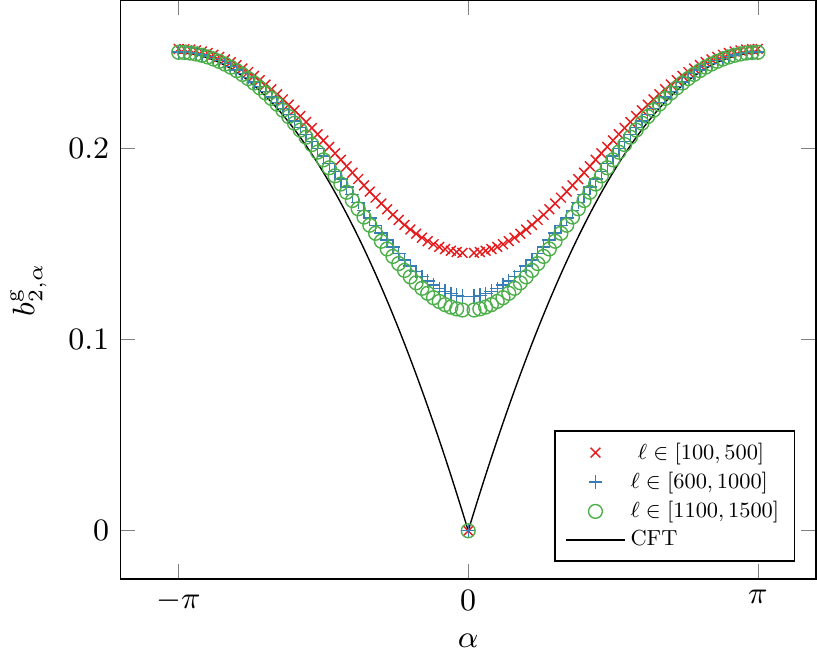}
    
     \caption{Numerical check of the CFT prediction~\eqref{eq:final_abs_charged_mom} for the absolute charged moments $Z_n^{\rm g}(\alpha)$ when $n=1$ (upper panel) and $n=2$ (lower panel). As explained
     in the main text, the points are the value of the coefficient
     $b_{n,\alpha}^{\rm g}$ in Eq.~\eqref{eq:delta_g_cft} obtained by 
     computing numerically $\Delta^{\rm g}$ in the non-Hermitian SSH 
     model for several lengths $\ell$ with $L = 2 \ell$ at fixed $\alpha$ and then fit Eq.~\eqref{eq:delta_g_cft} to them. The different symbols
     correspond to the three sets of subsystem lengths taken to 
     perform the fits: $\ell \in [100, 500]$, $\ell \in [600,1000]$, $\ell \in [1100,1500]$, in steps of $100$. The parameters of the Hamiltonian are $w=3/2$, $v=1$, $u=1/2$. The solid line is the CFT prediction~\eqref{eq:log_coeff_cft} for $b_{n, \alpha}^{\rm g}$.  }
     \label{fig:coeff-log-fit}
\end{figure}

We check numerically the result \eqref{eq:final_abs_charged_mom} in the following way. Let us define
\begin{equation}
	\Delta^{\rm g} \coloneqq \log\frac{Z_n^{\rm g}(\alpha)}{Z_n^{\rm g}(0)}-i\alpha \langle Q_A\rangle.
\end{equation}
According to  Eq.~\eqref{eq:final_charged_mom_ssh},  we expect
\begin{equation}\label{eq:delta_g_cft}
	\Delta^{\rm g}\sim b_{n,\alpha}^{\rm g}\log\ell
\end{equation}
for large $\ell$, with
\begin{equation}\label{eq:log_coeff_cft}
	b_{n,\alpha}^{\rm g}=-\frac{|\alpha|}{\pi n} \left(\frac{|\alpha|}{2\pi} -1 \right).
\end{equation}
The quantity $\Delta^{\rm g}$ can be exactly computed numerically for the 
ground state of the critical non-Hermitian SSH model using \cref{eq:abs_charged_mom_corr_matrix}.
Hence to verify that the coefficient $b_{n,\alpha}^{\rm g}$ has the form of \cref{eq:log_coeff_cft},  
we fit the curve $b_{n,\alpha}^{\rm g} \log\ell_{\rm e}$ to a set of numerical values of 
$\Delta^{\rm g}$ corresponding to different subsystem sizes $\ell=L/2$ with $\alpha$ 
and $n$ fixed. In Fig.~\ref{fig:coeff-log-fit}, we plot the values obtained in the fit for the 
coefficient $b_{n,\alpha}^{\rm g}$ in terms of $\alpha$, and compare it with the CFT 
prediction~\eqref{eq:log_coeff_cft}. To perform the fits, we consider three different sets of 
subsystem lengths, $\ell \in [100, 500]$, $\ell \in [600,1000]$ and $\ell \in [1100,1500]$ 
in steps of $100$. For $\alpha$ large enough, there is an excellent agreement 
between the CFT prediction and the result of the fit. This agreement is worse as 
$\alpha$ decreases, while at $\alpha =0$ there is again a very good matching.   
Repeating the fitting procedure with each set of points, we find that the larger $\ell$ the 
better the agreement with the analytic result. This behaviour suggests that the convergence
to \cref{eq:final_abs_charged_mom} for large $\ell$ is not uniform in $\alpha$.

Unfortunately, the non-universal factor $c_{n,\alpha}$ cannot be determined 
from the CFT or applying the usual lattice techniques employed in the case of the 
Hermitian SSH model, i.e. corner transfer matrix, block Toeplitz determinants.
Nevertheless, the numerical analysis of $c_{n,\alpha}$ reveals that the dominant
term is due to the momentum shift $\kappa$, which has a simple form
\begin{equation}\label{eq:non_univ_ct}
    c_{n,\alpha}\sim 
    \begin{cases}
          \kappa^{1-n} & \alpha=0, \\
          \kappa^{-n} & \alpha \neq 0.
    \end{cases} 
\end{equation}

\section{Symmetry-resolved entanglement entropy of the $bc$-ghost CFT}\label{sec:symm_res_ent_bc}

In this section, we use the results obtained in Eq.~\eqref{eq:final_abs_charged_mom} for the 
absolute charged moments $Z_n^{\rm g}(\alpha)$ to 
calculate the symmetry-resolved entanglement entropy 
$S_n^{\rm g}(\rhoaq)$.

To this end, according to \cref{eq:def-Sgnq}, we 
need to determine the Fourier transform of $Z_n^{\rm g}(\alpha)$. 
If we insert the CFT prediction of \cref{eq:final_abs_charged_mom} 
for the absolute charged moments in \cref{eq:f_trans_abs_charged_mom}, dropping the 
unknown non-universal term $c_{n,\alpha}$, then we obtain 
\begin{multline}\label{eq:symm_res_abs_mom}
    \Zg{n}(q) = (-1)^{\Delta q}Z_n(0)  \sqrt{\frac{n \pi\ell^{\frac{1}{n}}}{2\log\ell}} 
     e^{-\frac{n \pi^2 \Delta q^2}{2 \log \ell}} \\
    \times \Re \Erf \left( \frac{\log \ell + i \pi n \Delta q}{\sqrt{2 n \log \ell}}\right), 
\end{multline}
with $Z_n (0) = \ell^{1/3(n - 1/n)}$ and $\Delta q = q - \Qav$. 

For large $\log \ell$, the error
function has the asymptotic form 
$\Erf(x)\sim 1-e^{-x^2}/(\sqrt{\pi}x)$
and, therefore, Eq.~\eqref{eq:symm_res_abs_mom} can be expanded as
\begin{multline}\label{eq:exp_symm_res_abs_mom}
    \Zg{n}(q)~\sim Z_n(0)\left[(-1)^{\Delta q}  \sqrt{\frac{n \pi \ell^{\frac{1}{n}}}{2\log\ell}} 
    e^{-\frac{n \pi^2 \Delta q^2}{2 \log \ell}}\right. \\
    \left. -\frac{n\log\ell}{(\log\ell)^2+(n\pi \Delta q)^2}\right].
\end{multline}
Observe the leading term in $\ell$ is a Gaussian with an alternating
sign $(-1)^{\Delta q}$, while the subleading one has Lorentzian
form. It is important to not neglect this term, because it is a correction only for fixed $|\Delta q|\ll \log\ell$, while it becomes leading in the regime $|\Delta q| \agt \log\ell$. To understand this crossover, let us focus on the  case $n=1$ which gives the normalization term 
$p(q)=\mathcal{Z}_1^{\rm g}(q)$ in the charge sector 
decomposition~\eqref{eq:charge_dec_rhoa} of $\rhoa$,
\begin{equation}\label{eq:p_q_bc}
    p(q)~\sim (-1)^{\Delta q} 
    \sqrt{\frac{\pi\ell}{2\log(\ell)}}   
    e^{-\frac{ \pi^2 \Delta q^2}{2 \log \ell}}
    -\frac{\log\ell}{(\log\ell)^2+(\pi q)^2}.
\end{equation}
Since $p(q) = \tr \left(\Pi_q \rhoa\right)$, this quantity
must satisfy $\sum_q p(q)=1$. To get this, we have actually to take
into account the two terms of Eq.~\eqref{eq:p_q_bc}. One can check that
\begin{equation}\label{eq:norm_p_check}
    \sum_{q\in \mathbb{Z}} p(q)=\sqrt{\frac{\pi\ell}{2\log\ell}} 
    \vartheta_4(e^{-\pi^2/(2\log \ell)})-\frac{\ell^2+1}{\ell^2-1},
\end{equation}
where the first term comes from the Gaussian  in Eq.~\eqref{eq:p_q_bc} and the second one from the Lorentzian.
Expanding for large $\ell$, which is the only regime where the above equation makes sense (using
  $\vartheta_4(e^{-x})\sim 2\sqrt{\pi}e^{-\pi^2/(4x)}/\sqrt{x}$ for $x\sim 0$), we find that Eq.~\eqref{eq:norm_p_check}
tends to one in the large $\ell$ limit, but the Gaussian alone would provide $2$ and $-1$ comes from the Lorentzian. 
It is then clear that both terms should be always taken into account. 

Another interesting property of Eq.~\eqref{eq:exp_symm_res_abs_mom} is that, for 
large $\ell$, the sign of $\mathcal{Z}_n^{\rm g}(q)$ is $(-1)^{\Delta q}$. In particular, in the case $n=1$, this means that the sum of the eigenvalues of $\rhoa$ in each charge sector 
of the $bc$-ghost theory has sign $(-1)^{\Delta q}$ (as long as $|\Delta q|\alt \log\ell$).
The presence of this alternating sign can be understood by the relation 
\begin{equation} \label{eq:rhoa=-1Q}
    \rhoa = (-1)^{Q_A - \langle Q_A\rangle} \abs \rhoa.
\end{equation}
This identity can be derived in the lattice for any subsystem of arbitrary length $\ell$. In fact, along the critical line $u=w-v$, half of the eigenvalues 
$\nu_j$ of two-point correlation matrix $C_A$ of \cref{eq:correl_matrix} 
are $\nu_j < 0$ and the other half $\nu_j > 1$. Given the relation \eqref{eq:corr_matrix_ent_ham} 
between the two-point correlation matrix and single particle entanglement Hamiltonian $H^{\rm E}$,
the eigenvalues $\varepsilon_j$ of the latter are complex
\begin{equation}\label{eq:eigenvalues_ent_ham_corr_matrix}
 \varepsilon_j = \log \left|\frac{1-\nu_j}{\nu_j}\right|+i\pi.
\end{equation}
We can construct the eigenvalues of $\rhoa$ from those of $H^{\rm E}$. 
If we label them with the occupation numbers relative to the single 
particle entanglement Hamiltonian, then 
\begin{equation}
\lambda_{\{ n_j\}} = 
\mathcal{N}^{-1} e^{- \sum_j \varepsilon_j n_j}.
\end{equation}
Since $\mathcal{N}=\prod_{j=1}^{2\ell}
(1-\nu_j)$ and half of the eigenvalues $\nu_j$ are $\nu_j<0$ and the other half $\nu_j>1$,
then $\mathcal{N}=(-1)^\ell|\mathcal{N}|$. Combining this with \cref{eq:eigenvalues_ent_ham_corr_matrix} and 
identifying $\sum_j n_j = q$ and $\langle Q_A\rangle = \ell$, we obtain
\begin{equation}\label{eq:eigen_density_mat}
\lambda_{\{n_j\}}=(-1)^{q-\langle Q_A\rangle} |\lambda_{\{n_j\}}|.
\end{equation}
from which \cref{eq:rhoa=-1Q} follows. 

This result indicates that the non-positiveness of $\rhoa$ is due
to the global sign $(-1)^{q-\langle Q_A\rangle}$ on each charge
sector $q$. Observe that in the decomposition of \cref{eq:charge_dec_rhoa}, this factor 
is absorbed in the normalization $p(q)$ and, therefore, the density matrices
$\rho_{A, q}$ are positive definite. This implies that the generalized entanglement
entropy and the standard entanglement entropy of $\rho_{A, q}$ are equal,
$S_n^{\rm g}(\rhoaq) = S_n(\rhoaq)$.

Plugging \cref{eq:exp_symm_res_abs_mom} 
into \eqref{eq:def-Sgnq}, we obtain that the symmetry-resolved R\'enyi entanglement 
entropy in each charge sector behaves when $\log \ell\gg 1$ as
\begin{equation}\label{eq:symm_res_ent_ssh}
    \Sg_n(\rhoaq) = \frac{n+1}{6n} \log \ell-\frac{1}{2} \log \log \ell+ O(\ell^0).
\end{equation}
and, in the limit $n\to 1$,
\begin{equation}\label{eq:vn_symm_res_ent_ssh}
    \Sg(\rhoaq) = \frac{1}{3} \log \ell- \frac{1}{2}\log \log \ell+ O(\ell^0).
\end{equation}
Note that, contrary to the generalized entanglement entropy of the full reduced 
density matrix $\rho_A$, the symmetry-resolved entanglement entropies are positive. 
Moreover, the expansions of Eqs.~\eqref{eq:symm_res_ent_ssh} and \eqref{eq:vn_symm_res_ent_ssh} 
coincide with the symmetry-resolved entanglement entropies of the $1+1$ free massless 
Dirac fermion~\cite{Goldstein_2018, mdgc-20}. An important property of Eqs.~\eqref{eq:symm_res_ent_ssh} and 
\eqref{eq:vn_symm_res_ent_ssh} is that they do not depend on the charge $q$: at leading order in $\ell$, the 
symmetry-resolved entropy is equally distributed among all the charge sectors. This 
feature is known as entanglement equipartition~\cite{Xavier_2017}. As usually happens 
in Hermitian systems, one should further analyze the subleading $O(\ell^0)$ terms of 
\cref{eq:symm_res_ent_ssh} in order to find corrections that explicitly depend on $q$.
Unfortunately, we lack the proper tools to determine the first term that breaks the equipartition.

An interesting consistency test consists in recovering the total generalized entanglement entropy plugging in the decomposition~\eqref{eq:symmetry-res-decomposition} in charge sectors the result of Eq.~\eqref{eq:exp_symm_res_abs_mom} for $\mathcal{Z}_n^{\rm g}(q)$.
Unfortunately, performing the sum analytically is complicated due to the the alternating sign $(-1)^{\Delta q}$ and to the interplay of the Gaussian and Lorentzian terms. Anyhow, we have checked numerically that the sum ~\eqref{eq:symmetry-res-decomposition} behaves as 
\begin{equation}
    S^{\rm g}(\rhoa)\sim-\frac{2}{3}\log\ell,
\end{equation}
which is the correct result for the generalized entanglement entropy (obtained also by plugging  $Z_n^{\rm g}(0)$ of Eq.~\eqref{eq:abs_neutral_mom_ssh} in Eq.~\eqref{eq:gen_renyi_ent} and taking the limit $n\to 1^+$).

\section{Charged moments and standard R\'enyi entropies}\label{sec:ordinary_renyi_entropy}
In the previous section, we obtained that along the critical line $u=w-v$, the 
matrices $\rho_A$ and $|\rho_A|$ are related by \cref{eq:rhoa=-1Q}. 
Here we apply this identity to understand the dependence~\eqref{eq:standard_moments_ssh} 
on the parity of $n\in\mathbb{Z}$ of the standard moments of $\rho_A$, that we numerically observed in Fig.~\ref{fig:neutral_moments}.

We proceed as follows. First, in the definition ~\eqref{eq:standard_charged_moments} of the standard charged moments $Z_n(\alpha)$, we split $\rhoa^n = \rhoa \rhoa^{n-1}$. Then we use \cref{eq:rhoa=-1Q} in the $\rhoa^{n-1}$ factor to relate $Z_n(\alpha)$ to the absolute charged moments $Z_n^{\rm g}(\alpha)$ (cf. Eq.~\eqref{eq:abs_charged_mom}), obtaining
\begin{equation}\label{eq:rel_abs_standard_charged_mom}
 Z_n(\alpha)=e^{-i\pi (n-1)\langle Q_A\rangle}Z_n^{\rm g}(\alpha+\pi(n-1)).
\end{equation}
This equality implies that
\begin{equation} \label{eq:charged_moments_from_absolute_charged_moments}
    Z_n(\alpha) = 
    \begin{cases}
        Z_n^\mathrm g (\alpha) & n \text{ odd}, \\
        (-1)^{\langle Q_A\rangle} Z_n^\mathrm g (\alpha + \pi) & n \text{ even}.
    \end{cases}
\end{equation}
Employing the analytic expression obtained in \cref{eq:final_abs_charged_mom} for $Z_n^{\rm g}(\alpha)$, 
we find that, for large subsystem lengths $\ell$,
\begin{equation}\label{eq:final_charged_mom_ssh}
 Z_n(\alpha)= \begin{cases}
        c_{n, \alpha} e^{i\alpha\langle Q_A\rangle} \ell^{\frac{1}{3}\left(n-\frac{1}{n}\right)
        -\frac{|\alpha|}{\pi n}\left(\frac{|\alpha|}{2\pi}-1\right)} & n \text{ odd}, \\
        c_{n, \alpha+\pi} e^{i\alpha\langle Q_A\rangle} \ell^{\frac{n}{3}+\frac{1}{6n}-\frac{\alpha^2}{2n\pi^2}} & n \text{ even}.
    \end{cases}
\end{equation}
Note that, for $\alpha=0$, this result leads to the expressions conjectured in \cref{eq:standard_moments_ssh}
for the neutral moments of $\rho_A$. 
In \cref{eq:abs_charged_mom_bc}, we write the absolute charged moments of 
$\rho_A$ in the $bc$-ghost CFT as a correlation function of the 
composite twist fields $\tau_{n,\alpha}$ and
$\tilde{\tau}_{n,-\alpha}$. In the light of Eq.~\eqref{eq:final_charged_mom_ssh}, 
the standard charged moments are also given for $n$ odd by the same correlator, while for $n$ even we have to perform a shift in the phase $\alpha\mapsto \alpha+\pi$,
\begin{equation}
Z_n(\alpha)=
\begin{cases}
\bra{0}\tau_{n,\alpha}(0)\tilde{\tau}_{n, -\alpha}(\ell)\ket{0} & n \text{ odd},\\
\bra{0}\tau_{n,\alpha+\pi}(0)\tilde{\tau}_{n, -\alpha-\pi}(\ell)\ket{0} & n \text{ even}.
\end{cases}
\end{equation}

It is interesting to comment that the sensitivity of $Z_n(\alpha)$ to the parity of the exponent $n$
resembles the result for the entanglement negativity in Hermitian systems. The negativity is an entanglement 
measure for mixed states that can be obtained from the moments of the partial transpose of a given reduced 
density matrix~\cite{cct-12}. As for our density matrix $\rhoa$, the partial transpose in Hermitian systems is in general 
non-positive definite and, in unitary CFTs, its moments also display a dependence on the parity of $n$, 
similar to \cref{eq:final_charged_mom_ssh}, although with a different power law in $\ell$~\cite{cct-12, cct-13}.

\begin{figure}
    \centering
    \includegraphics[width=\columnwidth]{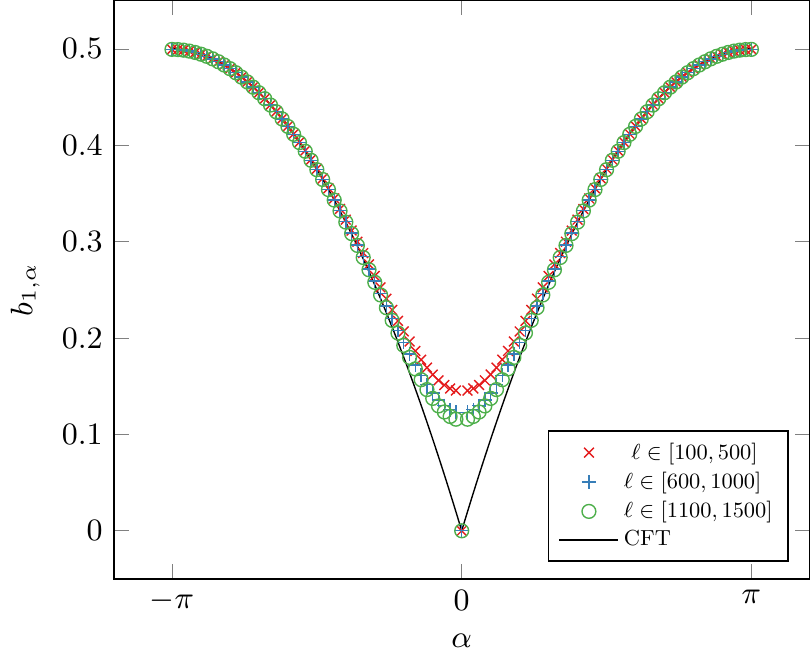}

         \hspace{0.2cm}
    \centering
    \includegraphics[width=\columnwidth]{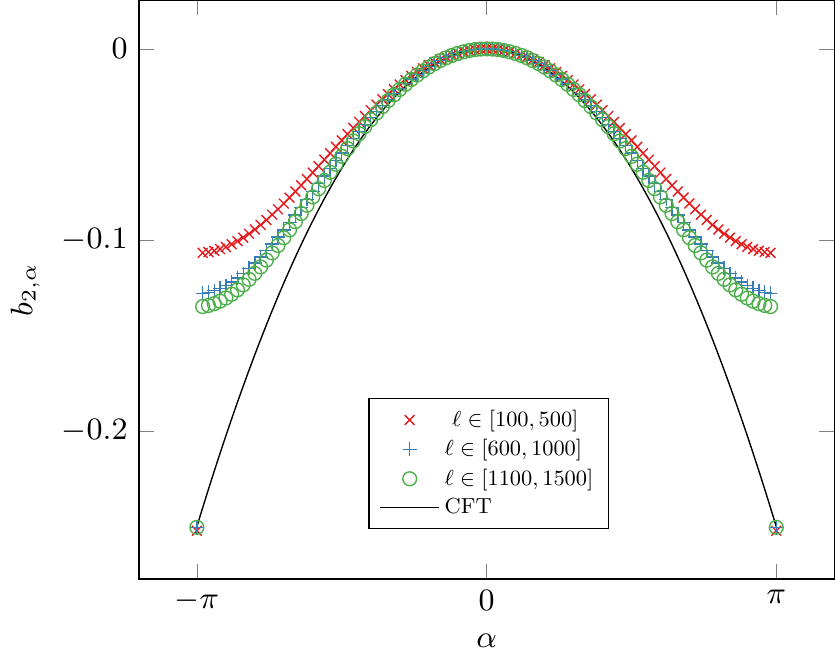}
    \caption{Numerical check of the analytic expression of Eq.~\eqref{eq:final_charged_mom_ssh}
    for the standard charged moments $Z_n(\alpha)$, with $n=1$ (upper
    panel) and $n=2$ (lower panel). As explained in the main text,
    the symbols are the results obtained for the coefficient $b_{n,\alpha}$ in a fit of Eq.~\eqref{eq:delta_cft} to a set of values of $\Delta$
    computed numerically in the non-Hermitian SSH model and corresponding to different subsystem lengths ($\ell\in[100, 500]$,
    $\ell\in[600,1000]$ and $\ell\in[1100, 1500]$ in steps of 100)
    for a given $\alpha$. 
    The parameters of the Hamiltonian are $w=3/2$, $v=1$, $u=1/2$. The solid curves are the analytic prediction of Eq.~\eqref{eq:pred_b_standard}  for $b_{n,\alpha}$ and large subsystem size. }
    \label{fig:num_check_standard_charge_mom}
\end{figure}

The results of Eq.~\eqref{eq:final_charged_mom_ssh} for the standard charged moments can be
checked with exact numerical calculations in the critical non-Hermitian SSH model by following the same strategy as for the absolute
charged moments presented in Sec.~\ref{sec:bc_ghost}. In fact, we consider the quantity
\begin{equation}\label{eq:delta_cft}
    \Delta \coloneqq \log\frac{Z_n(\alpha)}{Z_n(0)}-i\alpha\langle Q_A\rangle,
\end{equation}
which according to Eq.~\eqref{eq:final_charged_mom_ssh}, should behave as
\begin{equation}
    \Delta\sim b_{n,\alpha}\log\ell
\end{equation}
with
\begin{equation}\label{eq:pred_b_standard}
    b_{n,\alpha}=
    \begin{cases}
        -\frac{|\alpha|}{\pi n}\left(\frac{|\alpha|}{2\pi}-1\right) & n \text{ odd},\\
        -\frac{\alpha^2}{2\pi n^2} & n \text{ even}.
    \end{cases}
\end{equation}
We can exactly calculate $\Delta$ in the non-Hermitian SSH model
using Eq.~\eqref{eq:standard_moments_corr_matrix}. We therefore obtain numerically its value for several
subsystem lengths, with $\alpha$ and $n$ fixed, and we fit the function 
$b_{n,\alpha}\log\ell_{\rm e}$ to them. In \cref{fig:num_check_standard_charge_mom}, we show
the outcome for the coefficient $b_{n,\alpha}$ in the fit as
a function of $\alpha$ and two particular values of $n$, $n=1$ (upper 
panel) and $n=2$ (lower panel). The different symbols in the 
plot correspond to 
different sets of subsystem lengths chosen to perform the fit. The 
continuous black curves are the analytic prediction of \cref{eq:pred_b_standard} for 
large $\ell$. Observe that, as we have found in \cref{eq:rel_abs_standard_charged_mom}, there is 
relative phase shift $\alpha\leftrightarrow \alpha+\pi$ between $n$ even and
odd, consequence of the form~\eqref{eq:rhoa=-1Q} of the 
reduced  density matrix. Similarly to the absolute charged moments, 
cf. \cref{fig:coeff-log-fit}, the figure shows that the convergence to the asymptotic 
expression is not uniform in  $\alpha$ and worsens when approaching the cusp.

The oscillation in $n$ of the moments of
$\rho_A$ implies that the standard R\'enyi entanglement entropies also present
such behaviour. From \cref{eq:final_charged_mom_ssh}, we straightforwardly get
\begin{equation}\label{eq:standard_renyi_ent_ent_ssh}
    S_n(\rhoa) = 
    \begin{cases}
        - \frac{1}{3} \left( 1 + \frac{1}{n} \right) \log \ell + O(\ell^0) & n \text{ odd}, \\
        \frac{1}{1-n} \left(\frac{n}{3} + \frac{1}{6n} \right) \log \ell + O(\ell^0) & n \text{ even}.
    \end{cases}
\end{equation}
Finally, note that if we apply in the definition~\eqref{eq:vn_ent_ent} of the 
standard von Neumann entanglement entropy the relation~\eqref{eq:rhoa=-1Q} 
between $\rhoa$ and $|\rhoa|$, and we take into account that 
$\langle Q_A\rangle =\tr(\rhoa Q_A)$, we find that it coincides in
this case with the generalized one,
\begin{equation}
    S(\rho_{A})=S^{\rm g}(\rho_{A})\sim -\frac{2}{3}\log\ell.
\end{equation}
This asymptotic behaviour can be also extracted by taking the limit
$n\to 1^+$ in the expression ~\eqref{eq:standard_renyi_ent_ent_ssh} of the standard R\'enyi entropies for $n$ odd. This result agrees with the one obtained
in Refs.~\cite{Chang_2020, narayan} for this quantity. However, to our knowledge, the alternating behaviour in $n$ given by Eq.~\eqref{eq:standard_renyi_ent_ent_ssh} of the standard R\'enyi entropies has not been reported in previous works.

\section{Entanglement spectrum}\label{sec:ent_spec}

Using the results obtained in the previous section for the 
moments of the reduced density matrix $\rhoa$, we can
investigate the distribution $P(\lambda)$ of the its 
eigenvalues, that is of the entanglement spectrum. This is 
defined as
\begin{equation} \label{eq:P(lambda)-def}
	P (\lambda) =  \sum_{j} \delta (\lambda - \lambda_j),
\end{equation}	
where $\lambda_j$ are the eigenvalues of $\rhoa$. In Ref.~\cite{CalabreseLefevre}, 
the distribution $P(\lambda)$ was determined for unitary CFTs from the knowledge 
of the moments $Z_n(0)$. That approach was extended to analyze the negativity 
spectrum, i.e. the eigenvalues of the partial transpose of a density matrix, in unitary 
CFTs~\cite{Ruggiero_2016} and free fermions~\cite{Ruggiero_2019}. As we already pointed
out, the moments of $\rhoa$ in our non-Hermitian system present the same qualitative
dependence on the parity of the exponent $n$ as the moments of the partial transpose
in unitary CFTs. We can therefore easily apply here the techniques of Ref.~\cite{Ruggiero_2016} to obtain 
$P(\lambda)$ in the ground state of the non-Hermitian SSH model at criticality, which is the entanglement spectrum of the $bc$-ghost CFT.

The main idea of Ref.~\cite{CalabreseLefevre}, and also \cite{Ruggiero_2016}, is that the distribution $P(\lambda)$ is univocally
determined by its moments 
\begin{equation}
Z_n(0)=\sum_{j}\lambda_j^n=\int d\lambda P(\lambda) \lambda^n. 
\end{equation}
In fact, the Stieltjes transform $f(s)$ of $\lambda P(\lambda)$, 
\begin{equation}\label{eq:stieljes_transform}
 f(s)=\frac{1}{\pi}\sum_{n=1}^{\infty}Z_n(0) s^{-n}=\frac{1}{\pi}\int d\lambda\frac{\lambda P(\lambda)}{s-\lambda},
\end{equation}
is an analytic function in the complex plane except along the support 
of $\lambda P(\lambda)$ on the real line, where it has a branch cut. 
The discontinuity of $\lambda P(\lambda)$ at the branch cut gives 
$P(\lambda)$ through the formula~\cite{Marcenko}
\begin{equation}\label{eq:dist_stieljes}
    P(\lambda)=\frac{1}{\lambda}\lim_{\epsilon\to 0^+}f(\lambda-i\epsilon).
\end{equation}
This result implies that there is a one-to-one correspondence between
the Stieltjes transform $f(s)$ and the distribution $P(\lambda)$. Therefore,
given that we know the form of the moments $Z_n(\alpha)$, the strategy is to compute
the function $f(s)$ as the Laurent series of \cref{eq:stieljes_transform} and then determine
$P(\lambda)$ by applying \cref{eq:dist_stieljes}.

Before proceeding, it is important two emphasize two points.
When checking numerically the analytic expressions that we obtain in the following, it is crucial to take into account the shift $\kappa$ in the momenta that we have to perform to regularize
the two-point correlation matrix~\eqref{eq:correl_matrix_2} of the lattice system. We recall that this shift enters in the expression~\eqref{eq:final_charged_mom_ssh} of the moments $Z_n(0)$ through the non-universal constant $c_{n,\alpha}$, whose dependence
on $\kappa$ was determined in Eq.~\eqref{eq:non_univ_ct}. The second relevant aspect is
to multiply the expressions~\eqref{eq:final_charged_mom_ssh} of the moments $Z_n(0)$ by a global
factor $g^{1-n}$, both for $n$ even and odd, that accounts for a possible global degeneracy of the entanglement spectrum in the non-Hermitian SSH model, as it is explained in detailed in Ref.~\cite{Alba_2017} for unitary CFTs.

Including those two extra ingredients, in Appendix~\ref{sec:appendix_dist}, we show in detail the calculation of the Stieljes transform $f(s)$,
whose final expression reads
\begin{multline}\label{eq:stieljes_transform_ssh}
	f(s)  = \frac{g \kappa}{2\pi} \sum_{k=0}^\infty \frac{(-\log\ell)^k}{3^k k!} \left[\operatorname{Li}_k \left(\frac{\lambda_M}{s}\right) - \operatorname{Li}_k\left(- \frac{\lambda_M}{s}\right) \right]  \\
	+  \frac{g}{\pi} \sum_{k=0}^\infty \frac{1}{k!} \left( \frac{\log\ell}{12} \right)^k \operatorname{Li}_k \left( \frac{\lambda_M^2}{s^2} \right),
 \end{multline}
where $\operatorname{Li}_k(z)$ stands for the polylogarithm function and 
\begin{equation}\label{eq:lambda_max}
 \lambda_M=\frac{\ell^\frac{1}{3}}{g\kappa}
\end{equation}
is the largest eigenvalue of $\rhoa$ for large subsystem size $\ell$. In fact, the limit $n\to\infty$ of the standard moments
$Z_n(0)$ gives the largest eigenvalue $\lambda_M$ of $\rhoa$ such that $\lim_{n\to\infty} Z_n(0)^{1/n}=\lambda_M$. If we take $n\to \infty$ in \cref{eq:final_charged_mom_ssh}, then we obtain that in our case $\lambda_M$ is precisely~\eqref{eq:lambda_max}.
Note that this limit does not depend on the parity on $n$ of the R\'enyi entanglement entropy in \cref{eq:final_charged_mom_ssh} and, therefore, it is well-defined.

If we plug now \cref{eq:stieljes_transform_ssh} in the inversion 
formula~\eqref{eq:dist_stieljes}, we can obtain the distribution 
$P(\lambda)$. This requires to apply some properties of the polylogarithm 
$\operatorname{Li}_k(z)$. The reader can find a thorough description 
of this calculation in Appendix~\ref{sec:appendix_dist}. The final 
expression for $P(\lambda)$ is
\begin{multline}\label{eq:dist_ent_spec_ssh}
    P(\lambda) = 
    g \frac{1+\kappa}{2} \delta(\lambda - \lambda_M) 
    + g\frac{1-\kappa}{2} \delta(\lambda + \lambda_M) \\
    + g \frac{\Theta(\lambda_M - |\lambda|) }{2\lambda \sqrt{\log (\lambda_M / |\lambda|)}}  
    \left[- \kappa \sqrt{\frac{\log \ell}{3}}J_1\left(2 \xi_\ell(\lambda)\right)
 \right. \\
 \left. + \operatorname{sgn}(\lambda)  \sqrt{\frac{\log \ell}{6}} I_1\left(\sqrt 2 \xi_\ell(\lambda) \right)\right],
\end{multline}
where $J_\nu(z)$ and $I_\nu(z)$ are the Bessel and modified Bessel 
functions of the first kind and
\begin{equation}\label{eq:scaling_variable}
	\xi_\ell(\lambda) = \sqrt{\frac{\log (\ell) \log(\lambda_M/\abs{\lambda})}{3}}.
\end{equation}

The entanglement spectrum distribution  \eqref{eq:dist_ent_spec_ssh} 
presents some remarkable properties. It is reminiscent of the negativity spectrum of two intervals \cite{Ruggiero_2016} rather than the one-interval entanglement spectrum\cite{CalabreseLefevre} of unitary CFTs. Its support 
is the interval $[-\lambda_M, \lambda_M]$. The delta peaks at $\lambda=\pm \lambda_M$
indicate that there is a finite contribution from these two eigenvalues 
to the (generalized) entanglement entropies. 
As the momentum shift $\kappa \to 0$, the contribution of the function $J_1$ vanishes and $P(\lambda)$ becomes an even function in $\lambda$.
Observe that the constant 
$g$ is still undetermined, we will fix it by comparing with the numerical data 
of the lattice model.

A non-trivial numerical check of the correctness of \cref{eq:dist_ent_spec_ssh} 
is to study the tail or number distribution function $n_\ell(\lambda)$, that is the mean 
number of eigenvalues larger than a given $\lambda$,
\begin{equation}
 n_\ell(\lambda)=\int_\lambda^{\lambda_M} d\lambda' P(\lambda').
\end{equation}
Inserting \cref{eq:dist_ent_spec_ssh} in this expression, 
we find for $\lambda>0$
\begin{equation} \label{eq:n(xi)}
    n_\ell(\lambda) = \frac{g\kappa}{2}   J_0(2 \xi_\ell(\lambda))
    +  \frac{g}{2} I_0(\sqrt 2 \xi_\ell(\lambda)).
\end{equation}
The main feature of this result is that the number distribution admits a particular form in
which the joint dependence on $\lambda$ and $\ell$ is combined through
the scaling variable $\xi_\ell(\lambda)$ such that
\begin{equation}\label{eq:scaling-variable-def}
    n_\ell(\lambda)=n(\xi_\ell(\lambda)),
\end{equation}
with $n(x)$ a universal function that does not depend on the subsystem
size. A similar property is found in the entanglement and negativity spectrum of unitary CFTs~\cite{CalabreseLefevre, Ruggiero_2016}. 

\begin{figure}
     \centering
     \includegraphics[width=\columnwidth]{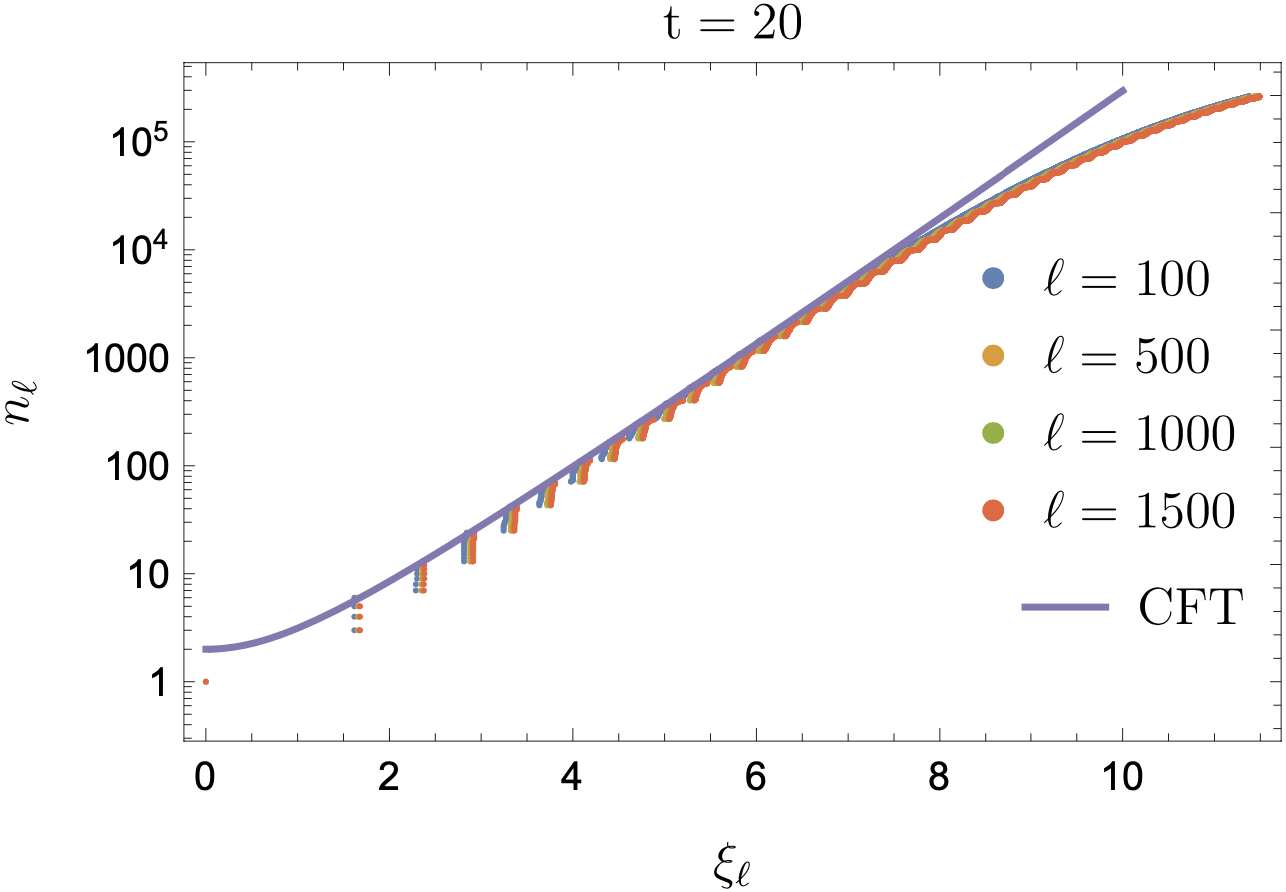}
     \includegraphics[width=\columnwidth]{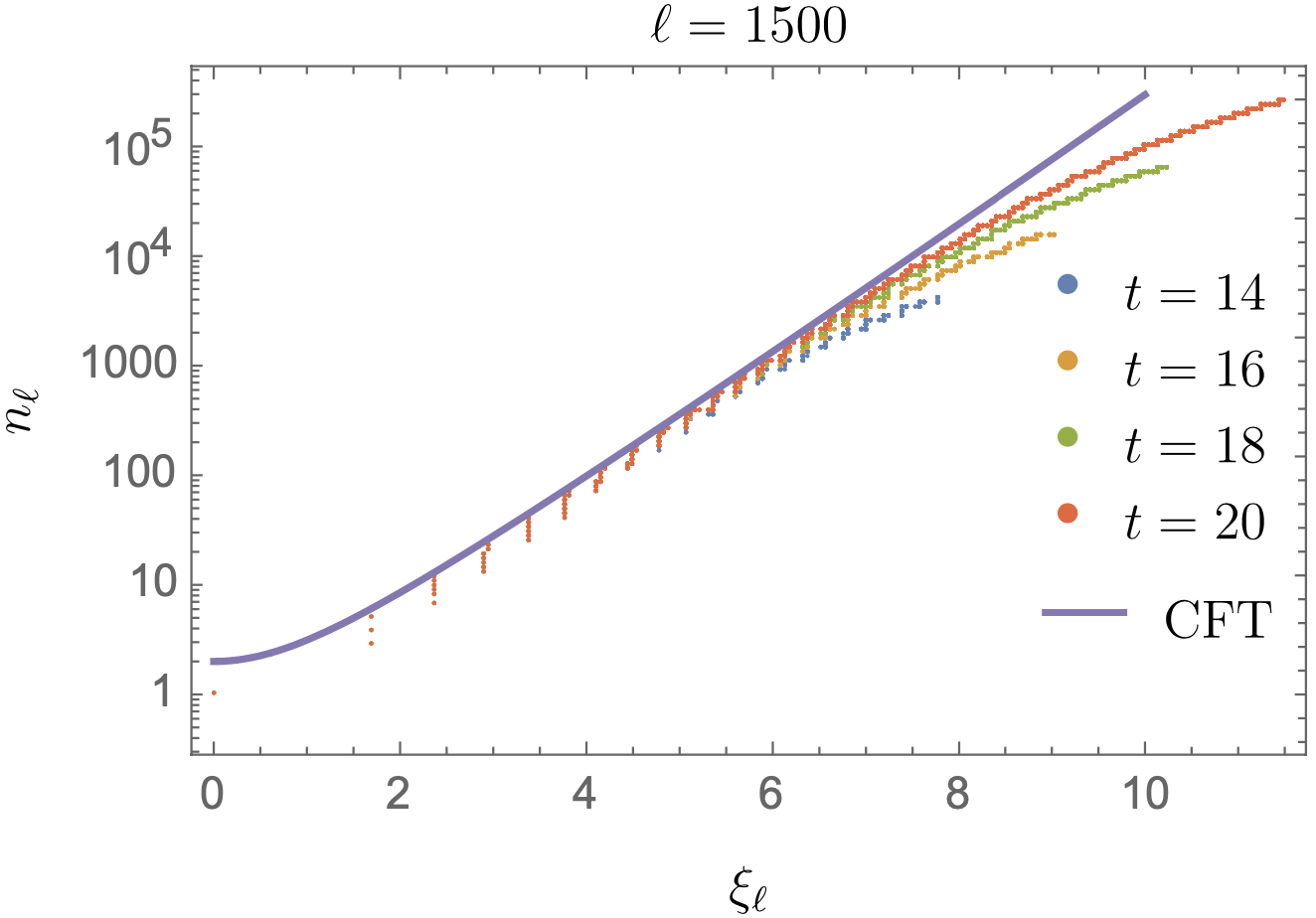}
     \caption{Number distribution of the entanglement spectrum 
     $n_\ell$ as a function of the scaling variable $\xi_\ell$ of
     \cref{eq:scaling_variable}. The points have been obtained 
     numerically as explained in the main text for the critical non-Hermitian SSH model and correspond to plot parametrically the 
     number of eigenvalues of $\rhoa$ in the interval $[\lambda, 
     \infty)$ in terms of $\xi_\ell(\lambda)$. In the upper panel, we 
     consider different subsystem sizes $\ell$, truncating the 
     spectrum of the correlation matrix to the $t=20$ largest 
     eigenvalues. In the 
     lower panel, we fix the subsystem size to $\ell=1500$ sites
     and we change the truncation to analyze its effect on the 
     numerical results. In both panels, the solid line represents the
     CFT prediction for $n(\xi)$ obtained in \cref{eq:n(xi),,eq:scaling-variable-def} 
     with $g=4$. The parameters of the Hamiltonian are $w=3/2$, $v=1$, $u=1/2$. }
    \label{fig:n-vs-xi}
\end{figure}

The structure of \cref{eq:scaling-variable-def} is particularly
useful when we want to make a comparison with the numerical data of the non-Hermitian SSH model. We can compute numerically the spectrum of $\rhoa$ for the non-Hermitian SSH model for a given subsystem length $\ell$ and plot, as we do in Fig.~\ref{fig:n-vs-xi}, the number
of eigenvalues $\lambda_j$ that lie in the interval $[\lambda, \infty)$ in terms of the scaling variable $\xi_\ell(\lambda)$ of
\cref{eq:scaling_variable}, taking as $\lambda_M$ the highest eigenvalue of the numerical spectrum and replacing $\ell$ by the chord length $\Le$.

If we repeat this procedure for different subsystem lengths, the numerical points should converge to the CFT
prediction for the number distribution of \cref{eq:n(xi)}. In
that expression, the only free term is the global factor $g$, which
is still undetermined. We numerically find that the spectrum of 
$\rho_A$ for the non-Hermitian SSH model presents a global two-fold
degeneracy. This forces, in particular, that the prefactor of the
delta peaks in $\lambda=\pm \lambda_M$ of the distribution~\eqref{eq:dist_ent_spec_ssh}
must be $2$ when $\kappa\to 0$. Thus we have to take $g=4$. In Fig.~\ref{fig:n-vs-xi}, the solid line corresponds to \eqref{eq:n(xi)} choosing this value for $g$. 
We obtain a very good agreement up to a certain value of $\xi_\ell(\lambda)$. In fact, for $\xi_\ell \to \infty$ ($\lambda\to 0$),
the CFT prediction for $n_\ell(\lambda)$ in Eq.~\eqref{eq:n(xi)} diverges 
since the number of eigenvalues in the continuum is infinite. The 
deviation of the numerical data for large $\xi_\ell$ is due to the 
finiteness of the entanglement spectrum on the lattice and the 
truncation method used to obtain it numerically.

The spectrum of $\rhoa$ for the non-Hermitian SSH model can be
determined numerically by means of the eigenvalues $\nu_j$ of the two-point correlation matrix, which is a $2\ell \times 2\ell$ matrix. 
Using \cref{eq:corr_matrix_ent_ham,,eq:eigenvalues_ent_ham_corr_matrix}, the $2^\ell$ eigenvalues $\lambda_j$ of the reduced 
density matrix are given by
\begin{equation}\label{eq:numerical_spectrum}
    \lambda_{\{n_j\}} = \prod_{j=1}^{2l} \nu_j^{n_j} (1-\nu_j)^{1-n_j},
\end{equation}
where they are labelled by a set $\{n_j\}$ of occupation numbers, cf. 
Eq.~\eqref{eq:eigen_density_mat}.
Since for large $\ell$ the storage of the $2^\ell$ eigenvalues $\lambda_{\{n_j\}}$ exhausts the memory capabilities, we restrict to the eigenvalues with largest absolute value and compute them with the following approximation. We truncate the spectrum of the correlation matrix to the first $t$ eigenvalues $\nu_j$ that maximize 
the distance with respect to $0$ and $1$ and we compute $\lambda_{\{n_j\}}$ according to \cref{eq:numerical_spectrum}, with $j=1, \dots, t$. The eigenvalues $\nu_j$ around $0$ or $1$ would produce a multiplicative factor that is near to either $0$ or to $1$. As long as we focus on the largest eigenvalues of $\rhoa$, like in Fig.~\ref{fig:n-vs-xi}, the multiplicative factor we are missing should be close to $1$ \cite{CalabreseLefevre}. In the upper panel of Fig.~\ref{fig:n-vs-xi}, we take a fixed truncation $t=20$ and consider
different subsystem sizes. In the lower panel, we study the effect of
the truncation for a given subsystem size; as clear from the plot, the distribution of the largest eigenvalues ($\xi_\ell\to 0$) is not affected when $t$ is increased.

\section{Conclusions}\label{sec:conclusions}

In this work, we have initiated the study of the symmetry resolution
of entanglement in non-Hermitian systems. In particular, we have 
considered the generalization of the R\'enyi entanglement entropy
based on the modulus $|\rhoa|$ of the reduced density matrix $\rhoa$ 
recently introduced in Ref.~\cite{Tu2022GenericEntropy}, which circumvents the problems that arise in the standard entanglement 
entropy due to the non-positiviness of $\rhoa$. Following an approach 
analogous to the Hermitian case~\cite{Goldstein_2018}, we have seen 
that the symmetry-resolved  entanglement entropy can be accessed 
through the Fourier transform of the moments of $|\rhoa|$, which we 
have called absolute charged moments. 

We have then focused on the ground state of the 
$bc$-ghost theory with central charge $c=-2$. This non-unitary CFT 
is the scaling limit of the non-Hermitian SSH model at criticality 
and has a global $U(1)$ symmetry. From a technical side, the main 
advantage of the non-Hermitian SSH model is that it is 
a quadratic fermionic chain, a fact that allows us to perform exact 
numerical calculations for large subsystem sizes thanks to Wick 
theorem. By applying bosonization techniques in the field theory, and 
with the support of exact lattice numerical calculations, we 
have derived the analytic expression of the absolute charged 
moments in the $bc$-ghost theory. They boil down to a two-point 
correlator of composite twist fields that properly includes the phase 
shift associated to the inserted charge.

From the result obtained for the absolute charged moments, we have 
found that the sign of the eigenvalues of $\rhoa$ in the $bc$-ghost 
CFT is determined by the charge sector to which they belong. This 
property is also true in the lattice system for any subsystem size.
Hence, since the eigenvalues of each charge sector have equal sign, 
we can define a positive-definite reduced density matrix in each 
sector and from it a positive symmetry-resolved entanglement entropy. 
Interestingly, we have obtained that the symmetry-resolved 
entanglement entropy of the $bc$-ghost is the same as the one of the 
massless Dirac fermion. 

We have also analytically determined the standard charged moments of 
$\rhoa$. They present different expressions when the R\'enyi index 
$n$ is either odd or even, a behaviour that stems from the dependence 
of the sign of the entanglement spectrum on the charge sector. This 
property is inherited by the standard R\'enyi entanglement entropies 
and resembles the case of the negativity in unitary 
CFTs~\cite{cct-12}. To our knowledge, this feature has not been 
reported in the literature, but it is expected to occur if the signs 
of the spectrum of $\rhoa$ have the same charge-resolved structure as 
in our case. We have seen that the standard charged moments of the 
$bc$-ghost theory are also given by a correlator of composite twist 
fields with a different phase shift for $n$ even or odd.

Finally, using the results for the standard moments of $\rhoa$, we 
have analytically derived the distribution of the entanglement 
spectrum in the $bc$-ghost theory, which is different from that 
of unitary CFTs~\cite{CalabreseLefevre}, although its number 
distribution turns out to be also a function of a scaling variable with no free parameters.

Our work leaves many open questions for future research. Here we have 
restricted to a particular non-unitary CFT. It would be interesting 
to study how our results extend to other theories or non-Hermitian
systems and, in particular, to see how universal are the expressions 
for the symmetry-resolved entanglement entropy and the entanglement 
spectrum distribution that we obtain for the $bc$-ghost theory with central charge $-2$.
In this respect, it would be needed a careful analysis using a path 
integral approach of the twist field correlators that give the 
(absolute) charged moments of $\rhoa$. 
Another relevant line would be to explore 
theories in which the spectrum of $\rhoa$ has other forms, 
e.g. it is complex, and wonder if the symmetry-resolved entropy is 
positive-definite as well. Moreover, here we have considered 
as ground state density matrix $\ket{\psi_R}\bra{\psi_L}$, but this 
is not the only possible choice; for example $\ket{\psi_R}
\bra{\psi_R}$, whose reduced density matrix is positive definite, is another reasonable alternative. It 
would be nice to study the symmetry-resolved entanglement entropies 
in such state and compare with our results.
Of course, the entanglement entropy is not the 
only entanglement measure that can be resolved in symmetry sectors, for example the 
negativity, which can be employed to study entanglement in disjoint 
subsystems~\cite{MBC-21, Chen-22}, or the operator entanglement \cite{Rath_2022}. One can also consider to carry out a similar analysis for them. 

\textit{Acknowledgements}. We thank L. Capizzi, A. Delmonte, J. Dubail, F. Essler, C. Muzzi, S. Murciano, F. Rottoli, C. Rylands, and S. Scopa for useful discussions. The authors acknowledge support from ERC under Consolidator grant number 771536 (NEMO).


\clearpage
\onecolumngrid
\appendix
\section{Details of the computation of $P(\lambda)$}\label{sec:appendix_dist}
In this appendix, we discuss in detail how to obtain the 
entanglement spectrum distribution $P(\lambda)$ of $\rhoa$
from its moments $Z_n(0)$ through the Stieljes transform $f(s)$.

In order to lighten the notation, let us rewrite the moments~\eqref{eq:final_charged_mom_ssh} 
of $\rhoa$ in the form
\begin{equation}\label{eq:mod_final_neutral_moments_ssh}
 Z_n(0)=\begin{cases}
   	    g \kappa e^{- bn + b'/n}, & n \text{ odd}, \\
     	g e^{- bn + b'/(2n)}, & n \text{ even},
    \end{cases}
\end{equation}
where
\begin{equation}
 e^{-b}=\frac{\ell^{1/3}}{g\kappa}, \quad
 e^{b'}=\ell^{1/3}.
\end{equation}
As explained in the main text, we have assumed in Eq.~\eqref{eq:mod_final_neutral_moments_ssh} that the non-universal term 
$c_{n,\alpha}$ is of the form of Eq.~\eqref{eq:non_univ_ct} and we have multiplied the expressions
both for $n$ even and odd by a global factor $g^{n-1}$ to take into account a possible global
degeneracy of the entanglement spectrum.

According to \cref{eq:stieljes_transform}, the moments of $\rhoa$ are the coefficients of 
the Laurent series expansion of $f(s)$. Taking int account the dependence
of $Z_n(0)$ on the parity of $n$, we split the series in the form 
\begin{equation}
 f(s)=\frac{1}{\pi}\left[\sum_{n=0}^\infty Z_{2n+1}(0)s^{-(2n+1)}+\sum_{n=1}^\infty Z_{2n}(0)s^{-2n}\right].
\end{equation}
Inserting \cref{eq:mod_final_neutral_moments_ssh} in this expression, and taking the Taylor series of $e^{b'/n}$ and
$e^{b'/(2n)}$ in the odd and even terms respectively,
\begin{eqnarray}
 f(s)&=& \frac{g}{\pi}\left[\kappa \sum_{n=0}^\infty e^{-b(2n+1) - \frac{b'}{2n+1}} s^{-(2n+1)} 
 + \sum_{n=1}^\infty e^{-b2n + \frac{b'}{4n}} s^{-2n}\right]\\
     &=&
 \frac{g}{\pi}\left[
  \frac{\kappa e^{-b}}{s} \sum_{k=0}^\infty \frac{(-b')^k}{k!} \sum_{n=0}^\infty \frac{(e^{-2b}/s^2)^n}{(2n+1)^k} 
    + \sum_{k=0}^\infty \frac{1}{k!} \left( \frac{b'}{4}\right)^k \sum_{n=1}^\infty \frac{(e^{-2b}/s^2)^n}{n^k}\right], 
\end{eqnarray}
we can then identify the Lerch trascendent 
$\phi(z,\nu,\alpha) = \sum_{n=0}^\infty \frac{z^n}{(n+\alpha)^\nu}$ 
in the odd term and the polylogarithm function ${\rm Li}_\nu(z)=\sum_{k=1}^\infty z^k/k^\nu$ in the even term,
\begin{equation}
    f(s)=\frac{g}{\pi}\left[\frac{\kappa e^{-b}}{s} \sum_{k=0}^\infty \frac{(-b')^k}{2^k k!} \mathrm \phi(e^{-2b}/s^2, k, 1/2) + 
    \sum_{k=0}^\infty \frac{1}{k!} \left( \frac{b'}{4}\right)^k \operatorname{Li}_k \left( \frac{e^{-2b}}{s^2}\right)\right].
\end{equation}
Finally, making use of the identity 
\begin{equation}
2^{-k} z \phi (z^2, k, 1/2) = \frac{1}{2} \left( \operatorname{Li}_k(z) - \operatorname{Li}_k(-z) \right),
\end{equation}
we arrive at \cref{eq:stieljes_transform_ssh}
\begin{equation}\label{eq:stieljes_transform_ssh_2}
f(s)= \frac{g\kappa}{2\pi} \sum_{k=0}^\infty \frac{(-b')^k}{k!} \left[ \operatorname{Li}_k \left(\frac{e^{-b}}{s}\right) - \operatorname{Li}_k\left(-\frac{e^{-b}}{s}\right) \right] + \frac{g}{\pi}\sum_{k=0}^\infty \frac{1}{k!} \left(\frac{b'}{4}\right)^k \operatorname{Li}_k \left( \frac{e^{-2b}}{s^2} \right).
\end{equation}

Once the Stieljes transform $f(s)$ is determined, we can obtain the 
distribution $P(\lambda)$ using the inversion formula~\eqref{eq:dist_stieljes}. 
The technical crucial point when applying \cref{eq:dist_stieljes} 
is that the polylogarith function ${\rm Li}_k(z)$ for $k\geq 1$ is a 
multivalued function with a branch point at $z=1$. If we take as 
branch cut the real interval $[1,\infty)$, the discontinuity across 
the cut is 
\begin{equation}\label{eq:polylog_jump}
   \lim_{\epsilon\to0^+} \Im \operatorname{Li}_k(y \pm  i \epsilon) =
    \begin{cases}
        \pm \pi (\log y)^{k-1}/\Gamma(k) & y \geq 1 \\
        0 & y<1.
    \end{cases}
\end{equation}
Using this property, we can easily obtain $P(\lambda)$ with 
Eqs.~\eqref{eq:dist_stieljes} and \eqref{eq:stieljes_transform_ssh_2}. 
To this end, let us consider separately the two terms of \cref{eq:stieljes_transform_ssh_2}. 
We define the function
\begin{equation}
f^{>}_\mathrm{odd}(s) = 
\frac{g \kappa}{2\pi} \sum_{k=1}^\infty \frac{(-b')^k}{k!} \left( \Li_k (e^{-b}/s) - \Li_k(- e^{-b}/s) \right),
\end{equation}
which corresponds to the first term of \cref{eq:stieljes_transform_ssh_2}, 
but removing the mode $k=0$. If we apply \cref{eq:polylog_jump}, then we find
\begin{eqnarray}\label{eq:im_fodd}
    \lim_{\epsilon\to 0^+}\Im f^>_\mathrm{odd} (\lambda - i \epsilon)
    &=& \frac{g\kappa\Theta(e^{-b}-\lambda)}{2 \log(e^{-b}/|\lambda|)} 
    \sum_{k=1}^\infty \frac{1}{k! \Gamma(k)} \left( -b' \log \frac{e^{-b}}{|\lambda|} \right)^k \nonumber \\
    &=& - \frac{g\kappa\Theta(e^{-b}-\lambda)}{2 \log(e^{-b}/|\lambda|)}
         \sqrt{b' \log\frac{e^{-b}}{|\lambda|}}\,  J_1\left(2 \sqrt{b'\log\frac{e^{-b}}{|\lambda|}} \right)
\end{eqnarray}
where $\Theta$ is the Heaviside step function and $ J_1$ is the Bessel function 
of the first kind. Note that for the last equality we have used that $b' > 0$

Analogously, we define
\begin{equation}\label{eq:feven}
    f^>_\mathrm{even} (s) = \frac{g}{\pi}  \sum_{k=1}^\infty \frac{1}{k!} \left( \frac{a}{2} \right)^k \operatorname{Li}_k \left( \frac{e^{-2b}}{s^2} \right),
\end{equation}
which is the second term of $f(s)$ in \cref{eq:stieljes_transform_ssh_2} 
without the mode $k=0$. If we again apply here \cref{eq:polylog_jump}, 
we have to take into account that for small $\epsilon$, the argument of the polylogarithm
in \cref{eq:feven} behaves as
\begin{equation}
\frac{e^{-2b}}{(\lambda - i \epsilon)^2} = \frac{e^{-2b}}{\lambda^2} + i \epsilon 2 \frac{e^{-2b}}{\lambda^3} + o(\epsilon^2)
\end{equation}
and, therefore, it approaches the branch cut from above or from below depending on the sign of $\lambda$. It then follows that
\begin{eqnarray}\label{eq:im_feven}
    \lim_{\epsilon\to 0^+}{\rm Im}f^>_\mathrm{even} (\lambda - i\epsilon) &=&
   \sign(\lambda) \frac{g \Theta(e^{-b}-|\lambda|)}{2 \log(e^{-b}/|\lambda|)} 
   \sum_{k=1}^\infty \frac{1}{k! \Gamma(k)} \left(\frac{b'}{2} \log \frac{e^{-b}}{|\lambda|} \right)^k \nonumber \\
    &=& g \Theta(e^{-b}-|\lambda|)\frac{\sign(\lambda)}{2 \log \frac{e^{-b}}{|\lambda|}}  
    \sqrt{\frac{b'}{2} \log \frac{e^{-b}}{|\lambda|}}\, I_1 \left( 2 \sqrt{\frac{b'}{2} \log \frac{e^{-b}}{|\lambda|}} \right)
\end{eqnarray}
where $I_1$ is the modified Bessel function of the first kind and in the last equality we have used $b' > 0$.

We still have to study the modes $k=0$ of $f(s)$ in \cref{eq:stieljes_transform_ssh_2},
\begin{equation}
    f_0(s) = \frac{g}{\pi} \Im \left[ \frac{\kappa}{2} \Li_0 \left( \frac{e^{-b}}{s} \right) 
    - \frac{\kappa}{2} \Li_0 \left( \frac{-e^{-b}}{s} \right)
    +  \Li_0 \left( \frac{e^{-2b}}{s^2} \right) \right].
\end{equation}
Taking into account that ${\rm Li}_0(z)=z/(1-z)$, they
can be rewritten as
\begin{equation}
    f_0(s) = \frac{g}{\pi}\left[\frac{\kappa}{2}\frac{1}{s/e^{-b}-1}+\frac{\kappa}{2}\frac{1}{s/e^{-b}+1}+\frac{1}{s^2/e^{-2b}-1}\right] .
\end{equation}
Applying the Sokhotski–Plemelj formula
\begin{equation}
 \lim_{\epsilon\to 0^+}\frac{1}{x\pm i \epsilon}=\mp i \pi \delta(x) + {\rm PV}\left(\frac{1}{x}\right),
\end{equation}
where ${\rm PV}$ stands for the Cauchy principal value, we get
\begin{equation}\label{eq:im_f0}
  \lim_{\epsilon\to 0^+}\Im f_0(\lambda - i \epsilon) 
  = g\lambda\frac{1+\kappa}{2} \delta(\lambda - e^{-b}) 
  + g\lambda \frac{1-\kappa}{2} \delta (\lambda + e^{-b}).
\end{equation}
Finally, putting Eqs.~\eqref{eq:im_fodd}, \eqref{eq:im_feven}, and \eqref{eq:im_f0} together 
in the inversion formula~\eqref{eq:dist_stieljes}, we get \cref{eq:dist_ent_spec_ssh}
\begin{multline}
    P(\lambda) = 
    g \frac{1+\kappa}{2} \delta(\lambda - \lambda_M) 
    +  g \frac{1-\kappa}{2} \delta(\lambda + \lambda_M) +\\
    + g
    \frac{\Theta(\lambda_M - |\lambda|) }{2\lambda \sqrt{\log (\lambda_M / |\lambda|)}}  
    \left[-\kappa \sqrt{b'} J_1\left(2 \sqrt{b' \log \left(\lambda_M /|\lambda|\right)}\right)
+  \operatorname{sgn}(\lambda)  \sqrt{b'/2} I_1\left(2 \sqrt{b'/2 \log \left(\lambda_M /|\lambda|\right)}\right)\right]
\end{multline}
where we have used that $e^{-b}= \lambda_M$.

\end{document}